\documentclass{article}
\usepackage{graphicx}
\usepackage{authblk}
\usepackage{tikz}
\usepackage{textcomp}
\usepackage{gensymb}
\usepackage{hyperref}
\usepackage{doi}
\usepackage{amsmath}
\usepackage{amssymb}
\usepackage{tablefootnote}
\usepackage{bbm}
\usepackage{subcaption}
\usepackage[section]{placeins}
\usepackage{float}
\usepackage[title]{appendix}
\usepackage{diagbox}
\usepackage{microtype}
\usepackage[margin=1in]{geometry}
\usetikzlibrary{positioning, calc}
\usetikzlibrary{shapes.geometric}
\usetikzlibrary{fit}
\usepackage[backend=biber,style=vancouver]{biblatex}
\begin{document}
\title{\Large An agent-based modelling approach to investigate the impact of gender on tuberculosis transmission in Uganda}
\author[1,*]{James W. G. Doran}
\author[2,3]{Dennis Mujuni}
\author[4]{Kit Gallagher}
\author[1]{Christian A. Yates}
\author[1]{Ruth Bowness}
%CHECK KIT AND DENNIS' DETAILS
\affil[1]{Centre for Mathematical Biology, Department of Mathematical Sciences, University of Bath, Bath, United Kingdom BA2 7AY}
\affil[2]{Department of Research and Development, Alpha Biomedical Research and Innovation Network, Kampala, Uganda}
\affil[3]{Makerere University, College of Health Sciences, Kampala, Uganda}
\affil[4]{Mathematical Institute, University of Oxford, Andrew Wiles Building, Radcliffe Observatory Quarter, Woodstock Road, Oxford OX2 6GG}
\affil[*]{Corresponding author: James W. G. Doran, jd521@bath.ac.uk}
\maketitle
\begin{abstract}
    Tuberculosis (TB) is an airborne disease caused by the pathogen \textit{Mycobacterium tuberculosis}. In 2023, it returned to being the leading cause of death from an infectious agent globally, replacing COVID-19; in the nineteenth century, one in seven of all humans died of tuberculosis. More than 10 million people are diagnosed with TB every year. The majority of cases in adults occur in males (62.5\% of all global adult cases in 2023, compared to 37.5\% in females). The main reasons for males suffering from a higher burden of global TB cases, compared to females, may be in large part due to population-scale factors, such as employment type, the quantity and type of social contacts they make, and their health-seeking behaviours (e.g. differences in diagnostic and treatment delays between genders). To investigate which population-scale factors are most important in determining this higher TB burden in males, we have developed an age- and gender-stratified, spatially heterogeneous epidemiological agent-based model. We have focused specifically on Kampala, the capital of Uganda, which is a high-burden TB country. We considered counterfactual scenarios to elucidate the impact of gender on the epidemiology of TB. Setting disease progression parameters equal between the genders leads to a reduction in both male-to-female case ratio and total case numbers.
\end{abstract}
\section*{Keywords}
%ADD KEY WORDS HERE - NEED 4-6
Tuberculosis, agent-based model, between-host dynamics, Python
%MAY REQUIRE HIGHLIGHTS SECTION (3-5 bullet points, up to 85 characters each) AS SEPARATE FILE
%\section*{Highlights}
%MAY REQUIRE GRAPHICAL ABSTRACT
%\section*{Graphical abstract}
\section{Introduction}
%FIRST PARAGRAPH - CONTEXT
Tuberculosis (TB) is an airborne disease caused by inhalation of the pathogen \textit{Mycobacterium tuberculosis} (\textit{M. tb}). According to the World Health Organisation, it ``probably returned'' to being the leading cause of death from an infectious agent globally in 2023, having been replaced by COVID-19 for the previous three years \cite{W.H.O.2024}. In the nineteenth century, one in seven of all humans died of tuberculosis \cite{Koch1982}; today, more than one million people die from the disease every year \cite{WHO2025}. It is estimated that approximately a quarter of the world's population are currently infected by the pathogen \cite{Houben2016}. In recent years, the majority of global incidence has occurred in high-burden TB countries: 86-90\% of the world's TB cases were distributed across thirty countries \cite{W.H.O.2024}. One of these countries is Uganda, which we have focused on in this study. Although efforts are being made to detect TB hotspots and increase case finding in Uganda, using data-driven AI tools like EPCON (which has been effectively used in other countries, such as Nigeria \cite{Alege2024}), it remains one of the worst affected nations, with an estimated 96,000 TB cases in 2023 \cite{WHO2024a}. Additionally, a greater number of TB cases in adults occur in males compared to females: in 2023, 62.5\% of adults who developed TB were males, whilst 37.5\% were females \cite{W.H.O.2024}.\par
The reasons for an increased TB burden in males compared to females could be in large part due to behavioural factors such as places of work, number of social contacts \cite{Nhamoyebonde2014} and sex assortativity in adult contacts, given that like-with-like social mixing has been found to be commonplace for both males and females \cite{Horton2020}. Uganda is one of the thirty high-burden TB countries; recent studies by \citeauthor{Nsawotebba2025} \cite{Nsawotebba2025} indicate that in the Ugandan capital, Kampala, on average, males experience a longer delay than females in diagnosis. Gender has been found to be a factor in delays in diagnosis and treatment in Uganda in other studies as well \cite{Buregyeya2014}, and it has been observed that men are less likely to receive a timely diagnosis compared to women in low- and middle-income countries in general \cite{Horton2016}. This may be related to different health-seeking patterns between males and females and would provide evidence for the `behavioural hypothesis' - one of two major hypotheses for the differences between genders in terms of  TB prevalence - due to males being exposed to other males who have not yet been diagnosed. This is in agreement with evidence that suggests males in Uganda with less knowledge of the symptoms of TB typically have longer diagnosis delays \cite{Obeagu2023}. The other major competing explanation is the `physiological hypothesis', which suggests differences in immune response and genetics are responsible for the increased prevalence of TB in males, although the two hypotheses are not necessarily mutually exclusive \cite{Nhamoyebonde2014}. Mathematical modelling is a popular and versatile method to investigate whether TB transmission in Kampala is significantly influenced by gender, and in particular whether the differences in diagnosis delay have an impact on the male-to-female ratios of TB cases and deaths from TB.\par
%SECOND PARAGRAPH - NEED
Modellers can investigate the spread of tuberculosis using a number of different mathematical approaches. Here we use an agent-based modelling framework. Agent-based models (ABMs) explicitly model each individual within the population of interest; each of these `agents' are assigned properties (such as age and gender), states (in the case of infectious disease models, these may include susceptible, infectious, recovered and others if appropriate) and rules to follow (who they are allowed to interact with, where to move if allowed - some examples could be moving up/down/left/right given a spatial location on a square lattice, or their location changing from `home' to `work' between time steps, and so on). Such a model simulates a disease outbreak by allowing these agents to interact with other agents and/or the environment, and transmit, or be infected by, the pathogen being investigated. Some previous epidemiological models of TB have used agent-based modelling frameworks (e.g. \cite{DeEspindola2011}, \cite{Guzzetta2011}, \cite{Tian2013}, \cite{Kasaie2015}, \cite{Prats2016}, \cite{Shrestha2017}, \cite{Ragonnet2019}, \cite{Zwick2021}, \cite{Udall2023}; see \cite{Bui2024} for a systematic review). Additionally, other TB transmission models have explored the impact of gender on the likelihood of developing active TB and dying from the disease (e.g. \cite{KisselevskayaBabinina2018}, \cite{Kubjane2023}, \cite{Wang2024}, \cite{Richards2025}). However, to our knowledge, no previous agent-based model has explored gender-differentiated TB transmission in a high-burden setting, such as Uganda, so our approach will fill a gap in existing research by combining gender/sex stratification with an agent-based modelling framework to study TB in a high-burden setting.\par
%THIRD PARAGRAPH - TASK
In this paper, we adapt an agent-based model, \texttt{Epiabm} \cite{Gallagher2024}, which in turn is an adaptation of \texttt{CovidSim}, one of the most influential models used to predict the progression of the COVID-19 epidemic within the United Kingdom and the impact of non-pharmaceutical interventions on this outbreak \cite{Ferguson2020}. \texttt{CovidSim} is an updated version of an earlier model of influenza pandemics \cite{Ferguson2005}, \cite{Ferguson2006}, \cite{Halloran2008}. The adapted version of \texttt{Epiabm} that we use throughout the rest of this paper will be referred to as \texttt{EpiabmTB}.\par
Firstly, we have adapted some of the features of the \texttt{Epiabm} framework to make it more specific to TB (for example, vaccination - from queued vaccination according to priority for COVID-19 to assuming the majority of individuals have already been vaccinated and determining whether newborns are vaccinated according to some probability for TB - and treatment using antimicrobials - from antivirals for COVID-19 to antibiotics for TB). Secondly, we have added in new elements (presented in Section \ref{sec:alterations}) to make the model more capable of capturing the epidemiological dynamics of TB that are not as relevant to COVID-19.\par
%FOURTH PARAGRAPH - OBJECT
We fit the model outlined in this paper to case count and death count data for a scaled-down version of the population of Kampala, Uganda to validate our approach, before using the model to explore counterfactual scenarios and observe their impact on the male-to-female ratio of TB cases and deaths. In this way, we can investigate the importance of different ``between-host'' factors (such as differences in diagnosis delays between genders, the proportions of each gender in schools/workplaces and sex assortativity of contacts) and ``within-host'' factors (such as differences in the probability of progression from latent TB to active TB and the probability of developing cavitary TB) in determining what contributes to the high TB burden observed in males, and therefore provide evidence for the behavioural hypothesis or the physiological hypothesis.\par
The rest of this paper is structured as follows. In Section \ref{sec:methodology}, we provide a brief overview of the alterations we have made to the \texttt{Epiabm} model in order to represent TB in Kampala with an age- and gender-stratified population. In Section \ref{sec:results}, we present the results generated using our model, comparing the calibrated data against counterfactual scenarios where differences between males and females are removed, as well as a consistency analysis. In Section \ref{sec:discussion}, we discuss the implications of our work and planned future work following on from this study.\par
\section{Methods}
\label{sec:methodology}
In this section, we give a brief overview of \texttt{CovidSim} and \texttt{Epiabm} - the previously developed agent-based models upon which our model is based - before discussing the addition, removal and alteration of features we have made to adapt the \texttt{Epiabm} model specifically towards modelling TB epidemiological dynamics, and then describing the parameter estimation for our model. In Appendix \ref{appendix:covidsim epiabm}, we provide a more detailed overview of \texttt{CovidSim} and \texttt{Epiabm}. More information on how \texttt{CovidSim} works can be found at \url{https://github.com/mrc-ide/covid-sim} and in \cite{Ferguson2020}. A full comparison of the two models and the differences between them can also be found at \url{https://github.com/SABS-R3-Epidemiology/epiabm}.
\subsection{\texttt{CovidSim} and \texttt{Epiabm}}
\label{sec:covidsim epiabm}
\texttt{CovidSim} is a spatial agent-based model designed to simulate infectious disease spread across a geographic region. Individuals are placed on a grid of cells subdivided into microcells, with households, schools, and workplaces allocated based on population density. Each person is assigned demographic attributes and an infection state drawn from standard compartments (Susceptible, Exposed, Infectious, Recovered, Removed), with infectious states further categorized by severity. Transmission occurs through household contacts, place-based interactions, and random social meetings, with probabilities influenced by distance via spatial kernels. The model initialises with most individuals susceptible and a small number infected, and infections propagate according to calculated forces of infection.\par
\texttt{Epiabm} is a modular reimplementation of \texttt{CovidSim}, offering Python and C++ backends and a more object-oriented structure for population generation and storage. While less computationally efficient, \texttt{Epiabm} is easier to adapt and extend. Both models support population initialisation from density files or random placement, assign household structures and ages using heuristic algorithms, and allocate schools and workplaces based on distance-weighted probabilities. Infection progression follows user-specified distributions, and transmission dynamics are updated at each time step, accounting for household, place, and inter-cell interactions. \texttt{Epiabm} simplifies some options (e.g. place size distributions) but retains core mechanisms from \texttt{CovidSim}.\par
We exploit the flexible framework provided by \texttt{Epiabm} for exploring epidemic dynamics in spatially structured populations, enabling detailed representation of demographic heterogeneity and contact patterns.
\subsection{Alterations}
\label{sec:alterations}
We have altered \texttt{Epiabm} to be more applicable to TB in a number of key aspects. We have redefined the disease states individuals can have at any point in time, as individuals can be carriers of latent TB infection for years after their initial exposure to \textit{M. tb} and can relapse even after successful treatment. In addition, we have added the ability of individuals to be born, age, die of non-TB causes, and move (i.e. be re-assigned to new) households and places; this evolution of the socio-demography has been added to take into account the fact that the serial interval is typically longer for TB relative to COVID-19. To investigate the role of gender on transmission dynamics, we have introduced gender as an attribute of individuals within the model, with separate proportions of each gender in each place type, the consideration of genders in household allocation, and separate diagnosis delay probability distributions.\par
Additional details about these alterations are given in Sections \ref{sec:disease states}-\ref{sec: more changes}, where we implement them and discuss the impact they have on the model results.
\subsubsection{Disease states}
\label{sec:disease states}
We have chosen the potential disease states an individual can have at any given point in time based upon those used by \citeauthor{Prats2016} \cite{Prats2016}. In total, there are 11 potential states. Newborn individuals initially enter the ``Healthy'' compartment (identical to the ``Susceptible'' compartment referred to in Appendix section \ref{sec:overview}).\par
If a healthy individual comes into contact with an individual currently experiencing active TB, there is the possibility that they are exposed to the \textit{M. tb} pathogen. If they are, they transition to the ``Latent TB'' compartment (this is identical to the ``Exposed'' compartment referred to in Appendix sections \ref{sec:overview} and \ref{sec:running the model}). If they recover naturally (that is, their immune response is able to completely remove all \textit{M. tb.} pathogens without the need for antibiotics), they will transition back to the healthy compartment after the period of time it takes the individual to recover naturally.\par
Alternatively, if individuals in the ``Latent TB'' compartment become sick, they transition to the ``Active TB'' compartment. This is identical to the ``Infectious'' compartment referred to in Appendix section \ref{sec:overview}, but is not sub-divided into categories based upon severity and the presence or absence of symptoms; it is assumed all individuals with active TB have equally severe disease and are equally symptomatic. It is worth noting that the Infectious compartment was sub-divided when compartmentalising COVID-19 infection states to help track hospitalisations and ICU visits more easily, which is not an aim of this paper, so the impact of not subdividing the infectious compartments on the model output should be negligible in this regard. Individuals in this compartment are capable of transmitting the disease to other healthy individuals. This can happen immediately after transitioning between the ``Healthy'' and ``Latent TB'' compartments (that is, the individual transitions from ``Healthy'' to ``Latent TB'' to ``Active TB'' in the same time step), according to the ``fast TB'' probability. Otherwise, it occurs after an exponentially distributed period of time, with this ``slow TB'' probability equal to one minus the fast TB probability.\par
Based on the TB model structure classification proposed by \citeauthor{Menzies2018} \cite{Menzies2018}, our model is closest to Structure E (implemented as Structure A with the possibility of transitioning from ``Healthy'' to ``Latent TB'' to ``Active TB'' in one time step - see Appendix \ref{appendix: latent to active TB} for an overview of each structure considered in \cite{Menzies2018}). \citeauthor{Menzies2018} found Structure E usually performed better than Structure A when replicating empirical data on TB transmission, although it underperformed compared to other structures (for instance, model structures that split the ``Latent TB'' compartment into a fast latent TB compartment and a slow latent TB compartment) \cite{Menzies2018}. However, due to limited data availability to inform parameter selection for these more complicated structures, we have chosen to implement a structure close to Structure E. It corrects for the issue present in Structure A which doesn't allow for an immediate decline in progression risk after infection. Unlike any of the structures explored in \cite{Menzies2018}, our model structure allows for natural recovery from infection and a return to the ``Healthy'' compartment, which should compensate for the fact Structure E typically overestimates long-term progression risk.\par
Infectiousness of individuals in the ``Active TB'' compartment is dependent on age; older individuals are more infectious than younger individuals \cite{Ragonnet2019}. In addition, with a certain probability, individuals can develop cavitary tuberculosis, with cavitations forming in their lungs, which increases infectiousness by a scalar multiplier. Males have a higher probability of developing cavitary TB than females \cite{Balogun2021}, which is reflected in the probabilities of having cavitary TB according to gender in our model. We have assumed that the infectiousness of individuals does not change over the course of their infection.\par
If the individual is diagnosed as having active TB, they transition to the ``Under Treatment'' compartment; they start receiving treatment and stop being infectious at this point. This is a valid assumption since evidence suggests effective treatment can reduce the transcription of genes associated with TB infectiousness after one day of treatment \cite{Shaikh2021}. Each gender has a separate diagnosis delay distribution when transitioning from the ``Active TB'' compartment to the ``Under Treatment'' compartment, reflecting the fact that gender appears to impact the delay between developing active TB and being diagnosed. From here, the individual receiving treatment may stop treatment, and subsequently relapse to the ``Active TB'' compartment; otherwise they will transition to the ``Recovered but Could Relapse'' compartment upon successful completion of their treatment.\par
\begin{figure}
    \centering
    \resizebox{\linewidth}{!}{
    \begin{tikzpicture}
        \node[draw, fill = green] (healthy) at (-3,0){\raggedright Healthy};
        \node[draw, text width = 1.5cm] (death healthy) at (-3,-3){\raggedright Non-TB death whilst healthy};
        \node[draw, fill = orange, text width = 1cm] (latent) at (-1,0){\raggedright Latent TB};
        \node[draw, text width = 1.5cm] (death latent) at (-1,-3){\raggedright Non-TB death whilst latently infected};
        \node[draw, fill = red, text width = 1cm] (sick) at (1,0){\raggedright Active TB};
        \node[draw, text width = 1.5cm] (death sick) at (1,-3){\raggedright Non-TB death whilst actively infected};
        \node[draw, fill = black, text = white] (dead) at (1, 2){\raggedright TB death};
        \node[draw, fill = blue, text = white, text width = 1.5cm] (treat) at (3,0){\raggedright Under\\treatment};
        \node[draw, text width = 1.5cm] (death treat) at (3,-3){\raggedright Non-TB death whilst under\\treatment};
        \node[draw, fill = yellow, text width = 1.6cm] (recover) at (5,0){\raggedright Recovered but Could Relapse};
        \node[draw, text width = 1.5cm] (death recover) at (5,-3){\raggedright Non-TB death whilst recovered};
        \draw[-stealth] (healthy.east) -- (latent.west);
        \draw ($(healthy.north)!0.75!(healthy.north east)$) |- (-2,1);
        \draw[-stealth] (-2,1) -| ($(sick.north west)!0.25!(sick.north)$);
        \draw[-stealth] (latent.west) -- (healthy.east);
        \draw[-stealth] (-5,0) -- (healthy.west);
        \draw[-stealth] (healthy.south) -- (death healthy.north);
        \draw[-stealth] (latent.east) -- (sick.west);
        \draw[-stealth] (latent.south) -- (death latent.north);
        \draw[-stealth] (sick.north) -- (dead.south);
        \draw[-stealth] (sick.east) -- (treat.west);
        \draw[-stealth] (sick.south) -- (death sick.north);
        \draw[-stealth] (treat.east) -- (recover.west);
        \draw[-stealth] (treat.west) -- (sick.east);
        \draw[-stealth] (treat.south) -- (death treat.north);
        \draw ($(recover.north)!0.75!(recover.north west)$) |- (3, 1);
        \draw[-stealth] (3, 1) -| ($(sick.north east)!0.25!(sick.north)$);
        \draw (recover.north) |- (1, 3);
        \draw[-stealth] (1,3) -| (healthy.north);
        \draw[-stealth] (recover.south) -- (death recover.north);
    \end{tikzpicture}
    }
    \caption{\raggedright Infectious disease states used in our TB epidemiological model and the potential transitions between them. The arrow into the Healthy compartment from the left of the figure represents births: newborns always enter the Healthy compartment at birth. See text for full description.}
    \label{fig:states}
\end{figure}
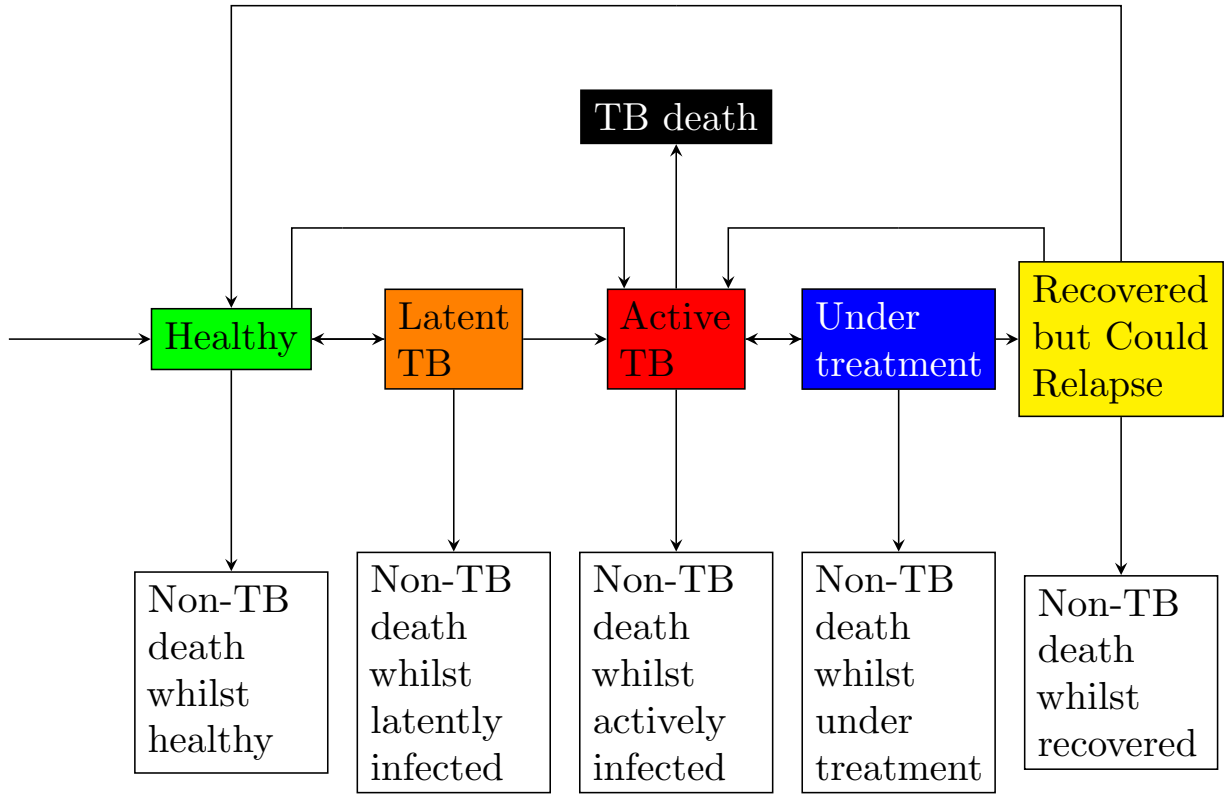

Individuals in the ``Recovered but Could Relapse'' state can relapse and return to the ``Active TB'' compartment, or transition back to the ``Healthy'' compartment once their risk of relapse is zero. For every one of the five non-newborn compartments previously mentioned, there is a compartment for those who die whilst in one of those states for reasons other than TB. Additionally, an individual experiencing active TB can die from TB and transition to the ``TB death'' compartment. We have assumed that once an individual starts treatment, their increased risk of dying from TB is removed, in keeping with \cite{Prats2016}. A schematic of the compartments and the transitions between them can be seen in Figure \ref{fig:states}.
\subsubsection{Evolution of socio-demography}
\label{sec:socio-demo evo}
The dynamics of TB infection play out over a longer period of time than those of COVID-19, which both \texttt{CovidSim} and \texttt{Epiabm} were originally designed to model. As such, we have added in the possibility for new individuals to be born, for individuals to age and potentially die of non-TB causes, and for individuals to move households and places. Our methodology is adapted from \cite{Guzzetta2011}. Births and non-TB deaths take place once per simulated day in our model, and ageing and relocation of households and places takes place once per simulated year in our model to reduce computational cost.\par
A new individual can be born into a household, provided the household has an adult female of suitable age already residing there who hasn't given birth within the previous nine months. For the parameter values used for age allocation within households, please refer to Table \ref{table:Reference parameter set JSON file} in Appendix \ref{appendix:parameter values}.\par
Individuals increase their age by one every year, and move into the next 5-year age group if old enough to do so. Individuals can either die from non-TB causes with a certain probability every day, or will do so automatically once they reach the maximum modelled age of 100, which we have assumed is the oldest age of individuals in the population based on data from \cite{UN2024}. This seems a reasonable assumption, as although approximately 1.1\% of the Ugandan population are 80 years old or older \cite{UBOS2024}, fewer than 0.01\% of Ugandans are 100 years old or older \cite{UN2024}. The date at which 100-year-olds are removed is chosen uniformly at random between the day they turn 100 and 364 days later, including both endpoints of this range, to increase the realism of the model. Due to insufficient data on social contacts for older individuals from \cite{Prem2017}, we have assumed all individuals aged 75 and older have the same contact rates.\par
There are two ways of moving household in our model: children becoming adults and moving into their own households, and married adults divorcing and moving into a single-person household. An individual that just became an adult and moves out from their original household, referred to as adult A for clarity, can partner up with somebody of a suitable age from the opposite gender; this couple then forms a new household in the same cell as adult A. We have made the assumption that in every household with a couple, the couple consists of one male and one female; this seems justified given that the Ugandan parliament passed an ``anti-homosexuality'' bill in 2023 \cite{Kakumba2023} and related laws formalised in the country since the 1902 Order in Council, when Uganda was a British protectorate \cite{Jjuuko2013}. Polygamy is legal and not uncommon in Uganda \cite{Mwanga2021}, so we have allowed for the possibility of husbands having a second wife in the same household. Whether or not a man has a second wife is determined by comparing the percentage of men with more than one wife to a number chosen uniformly at random between 0 and 1. We have assumed no man has more than two wives. In the case of divorce, one of the divorcees relocates to a new household within the same cell as the original household.\par
We have considered two place types in our model: schools and workplaces. This is different to both \texttt{CovidSim} and \texttt{Epiabm}, which split schools into primary, secondary and sixth form. The reduction in the number of types of schools is justified by the fact that children are rarely the source of tuberculosis transmission \cite{Logitharajah2008}. We have also removed care homes and outdoor spaces as place types; we have taken this step to simplify the model. Within each place type, we have two age groups, each with an associated proportion of the population belonging to that age group and gender which attend that place type (e.g. pupils and teachers in school). Once an individual ages beyond the maximum age of the age group they previously belonged to for a certain place type (e.g. once a pupil in school becomes too old to remain a student), they are removed from that place. They may then be assigned somewhere new, based on the proportion of individuals of their age and gender for the place type of the new place being proposed.
\subsubsection{Gender stratification}
\label{sec:gender strat}
Each individual is assigned a gender with probability drawn from a probability distribution determined according to gender distribution data in Uganda (see Table \ref{table:Reference parameter set JSON file} in Appendix \ref{appendix:parameter values}). In this model, we have assumed individuals are assigned one of two sexes at birth and identify as belonging to one of two genders. In addition, we have assumed each individual's gender identity aligns with their assigned biological sex, that is, individuals assigned the male sex at birth identify as men/boys and individuals assigned the female sex at birth identify as women/girls. Gender is considered along with age when distributing the population between the microcells, and when assigning individuals to places, during initialisation of the model: pupil-age females are more likely to be in school than pupil-age males, but adult-age males are more likely to be in workplaces than adult-age females, thereby capturing differences in workplace participation.\par
\begin{figure}
    \centering
    \begin{subfigure}[]{0.45\linewidth}
        \centering
        \includegraphics[width=\linewidth]{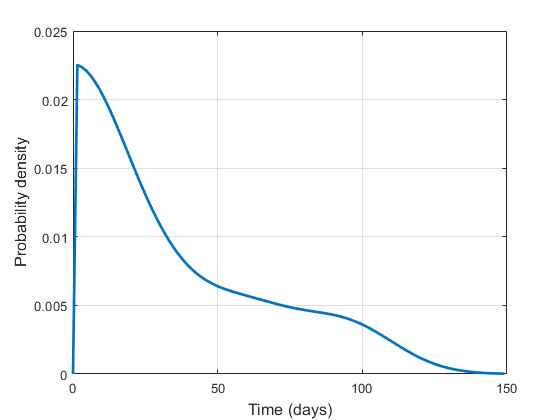}
        \phantomsubcaption
        \label{fig:TB-female-diag-delay}        
    \end{subfigure}
    \hfill
    \begin{subfigure}[]{0.45\linewidth}
        \centering
        \includegraphics[width=\linewidth]{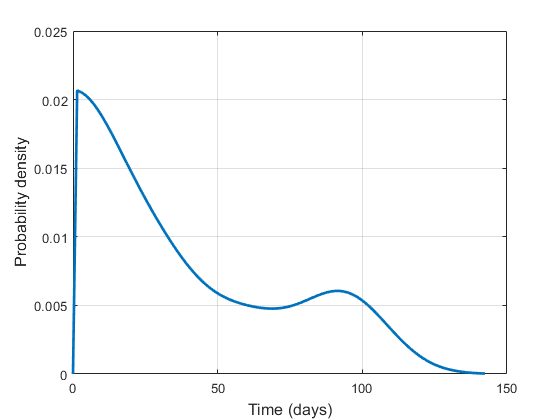}
        \phantomsubcaption
        \label{fig:TB-male-diag-delay}        
    \end{subfigure}
    \hfill
    \caption[]{Tuberculosis diagnosis delay distributions for females and males from Kampala, Uganda. The distributions are kernel density estimates with standard normal kernel functions and positive support using data from \citeauthor{Nsawotebba2025} \cite{Nsawotebba2025}}
    \label{fig:diag delay dist}
\end{figure}

The probability of transmission between a potential infector-infectee pair has been made partially dependent upon gender. Males are more likely to be infected than females, and transmission between individuals of the same gender is more likely than transmission between individuals of differing genders due to differences in daily contact rates resulting from increased sex assortativity of contacts \cite{Miller2021}. It is worth repeating in this section that an individual's diagnosis delay when in the Active TB compartment is determined according to their gender, with males having a longer average delay than females: this may be due to differences in treatment seeking behaviour. The diagnosis delay distributions for females and males are shown in Figure \ref{fig:diag delay dist}. In addition, the probabilities of transitioning from the Latent TB compartment to the Active TB compartment and from the Active TB compartment to the TB death compartment are scaled according to assigned sex (which we have assumed is aligned with gender identity, as noted above), to reflect the differences in disease progression rates; specifically the fact that males are more likely to become sick from, and die from, TB compared to females \cite{Gao2017} \cite{JimenezCorona2006} \cite{Fox2016}. Finally, it is worth repeating that we have allowed individuals with active TB to have cavitary disease: individuals with cavitation have a higher transmission rate \cite{Urbanowski2020} \cite{Balogun2021}. Evidence suggests males are more likely to develop cavitary tuberculosis \cite{Balogun2021}.
\subsubsection{Place size distributions}
\label{sec:place size dists}
We have changed the code so that workplace sizes are drawn from log-normal distributions instead of power-law distributions, due to insufficient information on Ugandan business sizes to be able to fit them to a power-law distribution. We have based our assumption that Ugandan business sizes are log-normally distributed on Gibrat's law of proportionate effect, which states that the growth rate of a business is independent of its current size \cite{Samuels1965}. A brief justification of why business size should be log-normally distributed if Gibrat's law holds is provided in Appendix \ref{appendix:lognormal business size}.\par
\begin{figure}
    \centering
    \includegraphics[width=\linewidth]{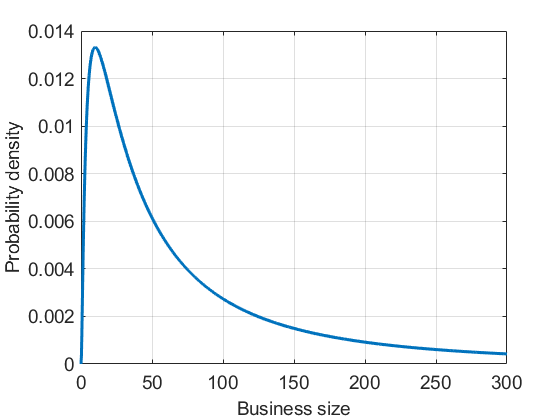}
    \caption{Log-normal distribution of Ugandan business sizes (by number of employees), with mean and standard deviation taken from \cite{Goyette2014}.}
    \label{fig:business size dist}
\end{figure}

Figure \ref{fig:business size dist} shows the assumed probability distribution for Ugandan business sizes used in our model.
\subsubsection{Additional alterations}
\label{sec: more changes}
To increase the speed of the initialisation of the model, in households with three or more individuals, the mean child age gap parameter was removed so that each child is assigned an age between the minimum and maximum child ages without deliberately spacing their ages. The model initialisation has also been modified to allow a user-specified number of individuals to belong to the ``Latent TB'', ``Active TB'', ``Under Treatment'' and ``Recovered but Could Relapse'' compartments. We have changed how the susceptibility of an individual is calculated by re-implementing the Who Acquired Infection From Whom (WAIFW) matrix from the \texttt{CovidSim} model to determine how susceptible an individual in one age group is to one from another. Additionally, we have added a WAIFW matrix stratified by gender, to account for research that suggests same-gender transmission appears to be more common than cross-gender transmission \cite{Miller2021}. We have altered how vaccination is implemented within the model, so that an initial proportion of all individuals is vaccinated and any newborn individuals are vaccinated with a certain probability thereafter. This is to reflect the fact that TB is an endemic disease in Uganda and most individuals are vaccinated by their first birthday, in contrast to COVID-19 which was a newly-emerging epidemic where the population had to be vaccinated during the initial outbreak with priority given to those more at risk.
\subsection{Parameter estimation, assumptions and limitations}
We have assumed that the transition times between the ``Latent TB'' and ``Active TB'' compartments (i.e. the latent period), between the ``Active TB'' and ``TB Death'' compartments (i.e. the infectious period), between the ``Under Treatment'' and ``Active TB'' compartments (i.e. the time to stopping treatment early), and between the ``Recovered but Could Relapse'' and ``Active TB'' compartments (i.e. the time to relapse) are exponentially distributed. For transitions where we had data on the cumulative probabilities of transitioning between states after certain lengths of time (e.g. from \cite{Esmail2014}, \cite{Borgdorff2011}), but not the mean length of time between transitions, we fit these distributions using the \texttt{fminsearch} function on \texttt{MATLAB} to find the parameter value that led to the best-fitting exponential distribution. The assumptions that the infectious period and time to relapse were exponentially distributed were based on work by \citeauthor{VanDenDriessche2007} \cite{VanDenDriessche2007}. The assumption that the latent period is exponentially distributed was also used for the model in \cite{CastilloChavez1997}, and it was later shown that replacing this assumed distribution with arbitrary latent period distributions led to similar results \cite{Feng2001}, which justify this approach that is simpler than some other epidemiological models of tuberculosis. The assumption that the time to stop treatment is exponentially distributed is equivalent to that used in \cite{Prats2016}, where individuals had a fixed daily probability of stopping treatment early, and whether they abandoned treatment or not was checked daily. Transition times between the ``Under Treatment'' and ``Recovered but Could Relapse'' compartments (i.e. the treatment period) are assumed to be constant: individuals transfer between the two compartments at the moment the full treatment period has been completed.\par
The diagnosis delay distributions for each gender were both derived using kernel density estimation with Gaussian kernels and positive support (see Figure \ref{fig:diag delay dist}). For households with six or more individuals (up to the maximum household size, stated in Table \ref{table:Reference parameter set JSON file} in Appendix \ref{appendix:parameter values}), we have assumed the proportions of households comprising a given number of people are Poisson-distributed with the mean equal to the average household size minus 1, based on research by \citeauthor{Jarosz2021} \cite{Jarosz2021}. That is, we start with a household occupied by one person, and the Poisson distribution determines the probability of a certain extra number of occupants living in that household. The entries in the WAIFW matrix are assumed to be proportional to contact rates between the age groups, based on research by \citeauthor{Wallinga2006} \cite{Wallinga2006}. We have assumed that all individuals are equally likely to transition between two given compartments regardless of age or gender. Estimates for the proportion of active TB cases with cavitation vary between 20\% and 90\% \cite{Gadkowski2008} \cite{Prats2016} \cite{Zhang2016} \cite{Urbanowski2020}, so we have assumed a proportion of 40\%. We have assumed that individuals with a cavitated form of the disease are 2.5 times as infectious, in keeping with the findings of a systematic review and meta-analysis of infectiousness risk factors for individuals with active TB \cite{Melsew2018}.\par
A full list of the numerical parameter values we used in our reference parameter set, and the justifications for using them, are given in Tables \ref{table:Reference parameter set workflow script} to \ref{tab:WAIFW gender} in Appendix \ref{appendix:parameter values}. It is worth noting that not all parameter values could be determined using Kampala-specific (or Uganda-specific) data.
\begin{figure}
    \centering
    \includegraphics[width=\linewidth]{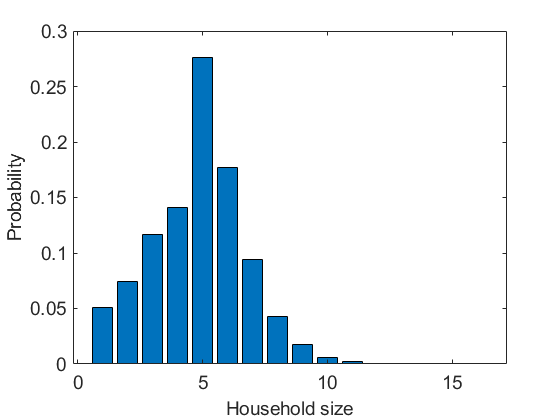}
    \caption{Assumed distribution of Ugandan household sizes, based on \cite{UBOS2024} and \cite{Jarosz2021}. Households of size 1 to 5 have probabilities taken from \cite{UBOS2024}; households of size 6 or more are calculated based on the method given in \cite{Jarosz2021}.}
    \label{fig:household dist}
\end{figure}
\section{Results}
\label{sec:results}
In this section, we present the outputs of the \texttt{EpiabmTB} model. Firstly, we confirm the model is able to replicate real-world data with calibrated parameter values. Secondly, we investigate the possibility of super-spreaders causing a majority of \textit{M. tb} transmissions in our model. Finally, we explore counterfactual scenarios to determine the relative impacts of certain physiological and behavioural factors in causing a high TB burden in males in Kampala. In addition, we conducted a consistency analysis to determine how many simulations were required to ensure outputs like averages and confidence intervals were consistent. For more details on the consistency analysis, please see Appendix \ref{sec:consistency analysis epiabm}.
\subsection{Calibrated results}
\label{sec:calibrated results}
We calibrated our model to ensure our model predictions were accurate by running a preliminary number of simulations and checking the total number of TB cases and male-to-female case ratio matched with what we expected given our data, tuning our parameter values until the averages and confidence intervals that were generated matched those that were observed in reality. After doing this, and subsequently completing the consistency analysis with the parameters found, we conducted 300 simulations of TB transmission in Kampala over a six-year time period, from 2017 to 2023. To reduce computational cost, we used a scaled-down population, in which one simulated individual was representative of 50 real individuals. Using a scaled-down population when simulating TB transmission using agent-based models is a commonly employed technique (see \cite{Bui2024} for examples) and there is evidence that using scaled-down populations can still lead to near-identical results to those obtained when using a full-scale population in agent-based models of disease transmission \cite{Hunter2022}. As such, the total number of simulated agents was 2\% of the population size of Kampala, Uganda (as estimated in 2019 \cite{UBOS2019}). It should be noted that the number of places (i.e. schools and workplaces) and households was also reduced to 2\% of the number of real places and households, so the number of infection events, case numbers, and amount of individuals in each compartment should match the expected amounts one would see for a population 2\% the size of Kampala's with the same demographic characteristics.\par
\begin{figure}
    \centering
    \begin{subfigure}{0.45\linewidth}
        \centering
        \includegraphics[width=\linewidth]{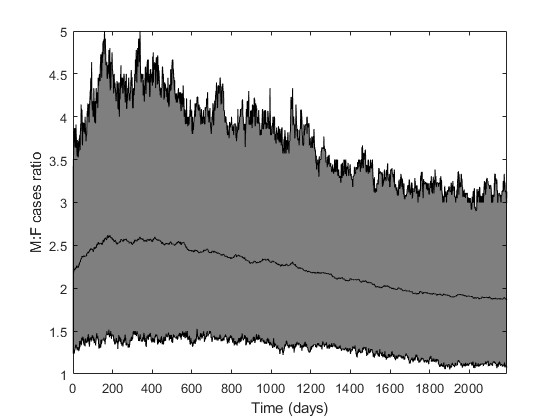}
        \label{fig:ratio results}
        \caption{Ratio of male-to-female TB cases.}
    \end{subfigure}
    \hfill
    \begin{subfigure}{0.45\linewidth}
        \centering
        \includegraphics[width=\linewidth]{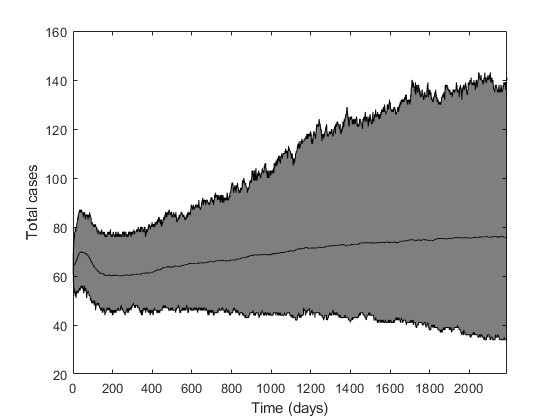}
        \label{fig:total cases results}
        \caption{Total number of TB cases.}
    \end{subfigure}
    \begin{subfigure}{0.45\linewidth}
        \centering
        \includegraphics[width=\linewidth]{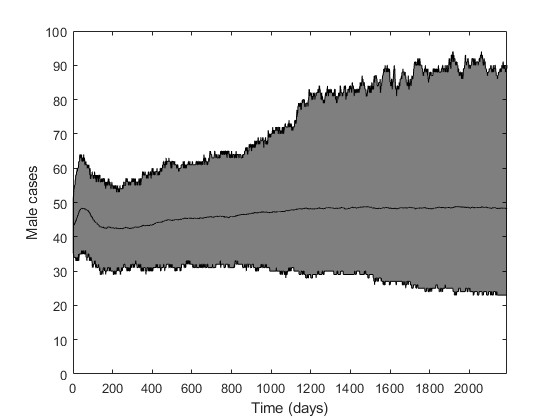}
        \caption{Male TB cases.}
    \end{subfigure}
    \hfill
    \begin{subfigure}{0.45\linewidth}
        \centering
        \includegraphics[width=\linewidth]{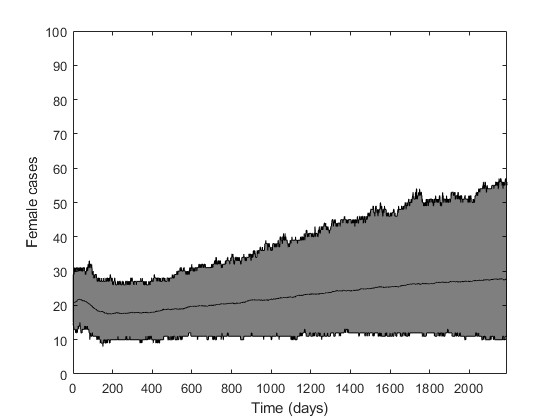}
        \caption{Female TB cases.}
    \end{subfigure}
    \caption[]{Summary plots of the male-to-female TB case ratio and total TB case numbers for all 300 simulations of the \texttt{EpiabmTB} model, as well as the TB cases stratified by gender. The shaded regions and the dashed lines at their boundaries indicate the 95\% confidence intervals. The solid lines contained within these regions indicate the mean amounts of the random variable in question.}
    \label{fig:results}
\end{figure}

A summary of the 300 calibrated simulation results is shown in Figure \ref{fig:results}. As can be seen, the number of TB cases grows each year on average, rising from the initial condition of 65 total cases to approximately 76 on average. However, the 95\% confidence intervals show that some simulations did see a decrease in the number of cases to fewer than 40 agents (34 was the lower bound of the confidence interval at the last time point of the simulations). Other simulations saw cases rise much more substantially over the time period simulated, with at least 140 total cases (the upper bound of the confidence interval for the last time point of the simulations was 140). This broadly matches the uncertainty in estimates of the real number of TB cases in Uganda over this time period: were the population of Uganda scaled down to be equal to the size of our simulated populations, the number of new TB cases per year from our simulations closely match the numbers that would be expected in reality \cite{WHO2024a}. The ratio of male-to-female TB cases is, on average, around 2 to 3 male cases for every female case for most of the simulated time period: the average ratio at the end of each simulation was 1.88 (to 2 decimal places). The 95\% confidence intervals show that, in some simulations, the ratio almost drops to 1 male case for every female case (the lower bound of the confidence interval at the last time point of the simulations was 1.09 to 2 decimal places). In other simulations, it rises to between 3 and 5 male cases for every female case (the upper bound of the confidence interval for the last time point of the simulations was 3.13 to 2 decimal places). Again, this is consistent with what has been observed in Uganda in reality \cite{Boum2014}.
\subsection{Super-spreaders}
One of our hypotheses for why the male-to-female TB cases ratio is potentially as high as 3-to-1 was the possibility of super-spreader events, with the high proportion of male-to-male contacts in particular, allowing a few male super-spreaders the opportunity to infect a large number of other males. Multiple previous studies have found that super-spreaders cause a large proportion of TB infections (see for example \cite{Ypma2013}, \cite{McCreesh2018}, \cite{Melsew2019}, \cite{Lee2020}, \cite{Teicher2023}, and \cite{Smith2023}). As males are more likely to suffer from cavitated TB \cite{Balogun2021}, and cavitated TB tends to make cases more infectious than non-cavitated TB \cite{Urbanowski2020}, the notion of males with cavitated TB being super-spreaders to other males seemed plausible.\par
To investigate this, whenever a transmission event occurred that led to infection, we tracked the gender of the infector responsible for transmitting the infection, as well as whether they had cavitation or not. Subsequently, we counted the number of infections each infector caused throughout a simulation and used this to fit a distribution of number of transmissions per infector (note that an infector had to have caused at least one infection to have been tracked).\par
In order to characterise the degree to which super-spreaders play on transmission dynamics, we will fit the number of transmissions per infector to a negative binomial distribution, $\text{NB}(r,p)$, taking inspiration from the methodology of \citeauthor{Goyal2022} (in \cite{Goyal2022}) (note that this probability distribution must be zero-truncated as we only count infectors who infected at least one individual). The negative binomial distribution is equivalent to a Poisson distribution, $\text{Pois}(\lambda)$, where $\lambda$ is a random variable distributed as a Gamma distribution, $\lambda \sim \Gamma(r,\frac{1-p}{p})$. We then set $r = \frac{\theta}{\rho}$ and $p = \frac{1}{1+\rho}$, where $\theta$ is equal to the mean number of transmissions per infector and $\rho$ is the index of dispersion, equal to the variance of the Gamma distribution divided by the mean. Thus, $\lambda \sim \Gamma(\frac{\theta}{\rho},\rho)$. A brief proof of the equivalence of the negative binomial distribution and the Poisson distribution with a Gamma-distributed rate parameter is given in Appendix \ref{appendix:NB gamma-Poisson equivalence}.\par
The negative binomial distribution is a convenient distribution to fit to the data, as the value of $\rho$ can directly indicate whether super-spreaders are causing a large proportion of infections. As the value of the shape parameter of a Gamma distribution increases (and thus the $r$ parameter in a negative binomial distribution increases), the Gamma distribution tends towards a normal distribution. For the same mean number of infections, $\theta$, a smaller index of dispersion (less than 1 and close to 0: $\rho \ll 1$) would lead to a larger shape parameter in our distribution ($\lim \limits_{\rho \to 0}\frac{\theta}{\rho} = \infty$), and would indicate that the number of infected individuals by a single infector is a random variable drawn from an approximately normal distribution with little variation. Conversely, a large index of dispersion greater than 1 ($\rho > 1$) would lead to a small shape parameter ($\lim \limits_{\rho \to \infty} \frac{\theta}{\rho} = 0$), and so indicate that the number of infections per infector is not approximately normally distributed. Furthermore, a small shape parameter leads to a more positively skewed Gamma distribution (and thus a more positively skewed negative binomial distribution), suggesting the distribution is right-tailed: in this instance, most infectors only infect a small number of individuals, whilst a small number of infectors cause a large number of infections (i.e. the mode number of infections is less than the median, which is less than the mean).\par
\begin{figure}
    \captionsetup{labelformat=empty,listformat=empty}
    \centering
    \begin{subfigure}{0.45\linewidth}
        \centering
        \includegraphics[width=\linewidth]{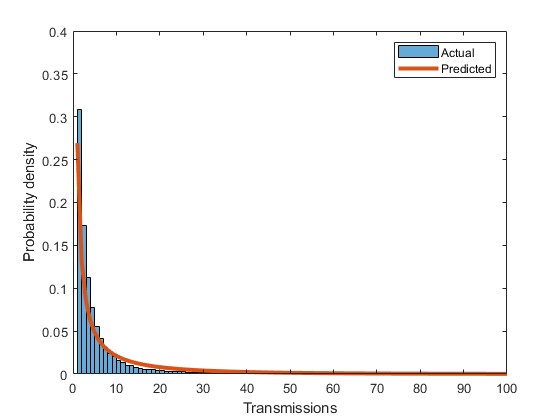}
        \caption{}
        \label{fig:superspreaders}
    \end{subfigure}
    \hfill
    \begin{subfigure}{0.45\linewidth}
        \centering
        \includegraphics[width=\linewidth]{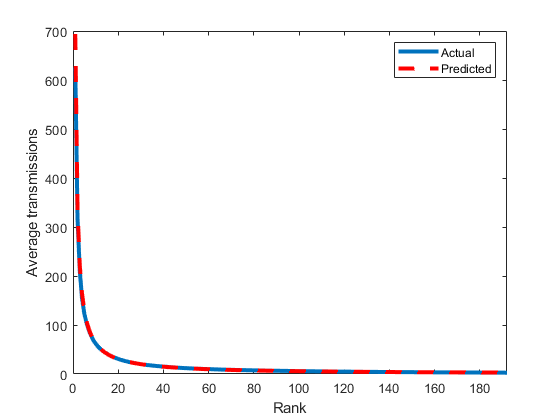}
        \caption{}
        \label{fig:transmission rank}
    \end{subfigure}
    \caption[]{Figure \ref{fig:super-spreaders}: (\subref{fig:superspreaders}) Probability distribution of the number of transmissions per infector (who infected at least one individual), averaged across the 300 simulations. The blue bars indicate the average probability density of the number of transmissions, and the orange curve shows the fitted zero-truncated negative binomial distribution; (\subref{fig:transmission rank}) Average number of transmissions caused by the biggest transmitters in any given simulation across the 300 simulations, and the best-fitting power-law approximation. The blue solid line shows the actual average, and the red dashed line shows the predicted average, calculated using the power-law approximation.}
    \label{fig:super-spreaders}
\end{figure}

Our results suggest that super-spreaders are responsible for a large proportion of TB infections in Kampala, Uganda. The zero-truncated negative binomial distribution fit to infections from all infectors (who caused at least one infection) across the 300 simulations has parameter values $r = 1.8850 \times 10^{-10}, p = 0.0269$ (to 4 decimal places), which, when viewed as a Poisson distribution with the rate distributed as a Gamma distribution with shape $r = \frac{\theta}{\rho}$ and scale $\rho = \frac{1-p}{p}$, implies $\theta = 6.8088 \times 10^{-9}, \rho = 36.1203$ (to 4 decimal places). The low value of $r$ (and consequently $\theta$) suggests the majority of individuals who are infected in a simulation do not transmit the disease to anybody else: either they have active TB for too short a period (before they die or are diagnosed) to transmit it to anyone else, or their infection remains latent. It should be noted that $\theta$ is much smaller than the sample mean, equal to 9.9932, as the distribution is a zero-truncated negative binomial, so the parameters are fitted to take the expected missing number of zero entries (i.e. those individuals who were infected and did not transmit TB, so were not recorded in the data set) into account. The distribution appears to provide a good fit to the data, with a squared correlation coefficient of 0.9884. The large value of $\rho$ implies a right-tailed distribution, as can be seen in Figure \ref{fig:superspreaders}, and suggests we can reject the hypothesis that the number of transmissions is approximately normally distributed. To confirm this, we conducted a one-sample Kolmogorov-Smirnov test on the data, which rejected the null hypothesis that the data is normally distributed at the 1\% significance level. This right-tailed distribution is also observed when we stratify infectors by gender, by cavitation status, and by both gender and cavitation status, as can be seen in Appendix \ref{appendix:stratified superspreaders}.\par
Furthermore, we analysed the proportion of transmissions caused by the top 20\% of infectors (hereafter referred to as $t_{20}$); on average, this came to 77.51\% (to 2 decimal places) across the 300 simulations, and ranged from 62.98\% to 90.10\%. This is largely in keeping with the 80/20 rule commonly observed in diseases with super-spreaders, where 20\% of infectious individuals cause approximately 80\% of infections \cite{Galvani2005}. This would imply the average number of transmissions per infector, when sorted by rank, should fit a power law function of the form $y = Cx^{m}$, where $y$ is the average number of transmissions and $x$ is the infector's rank. We found the best-fitting power law function had parameter values $C = 694.2174, m=-1.0417$ (to 4 decimal places), with a squared correlation coefficient of 0.9992 (see Figure \ref{fig:transmission rank} for the observed average number of transmissions per rank and the expected values according to the power law function). This is very close to satisfying Zipf's law: sorting the infectors in decreasing order of the number of transmissions they caused, the number of transmissions of the $n^{\text{th}}$ ranked infector is approximately inversely proportional to $n$.\par
Our data also indicated that males were causing a larger average number of transmissions than females (2,688 transmissions per simulation caused by males versus 1,185 caused by females, rounded to the nearest integer) and individuals with cavitated TB caused more transmissions than those with non-cavitated TB (2,745 versus 1,128 transmissions for cavitated versus non-cavitated TB, rounded to the nearest integer). Nevertheless, we wanted to find more definitive evidence that super-spreaders were more likely to be male and have cavitated TB, so we stratified the data by gender, by cavitation status and by both gender and cavitation status, and reran the analysis. The fitted parameter values for $\rho$ can be seen in Table \ref{tab:xbar rho}, along with the sample means.\par
\begin{table}
    \centering
    \begin{tabular}{|l|l|l|l|}
         \hline        
          $(\bar{x},\rho)$ & \textbf{Male} & \textbf{Female} & \textbf{Both}\\
         \hline
         \textbf{Cavitated} & (15.4882, & (13.1421, & (14.7915,\\
          & 64.8355) & 52.2314) & 61.0581)\\
          \hline
         \textbf{Not cavitated} & (6.0491, & (5.0275, & (5.5857,\\
          & 17.7214) & 13.4148) & 15.7373)\\
         \hline
         \textbf{Both} & (11.1621, & (8.0748, & (9.9932,\\
          & 41.9759) & 26.8695) & 36.1203)\\
          \hline         
    \end{tabular}
    \caption{Sample means, $\bar{x}$, and fitted parameter values for $\rho$ in the stratified zero-truncated negative binomial distributions with $r = \frac{\theta}{\rho},p=\frac{1}{1+\rho}$. As the distribution is zero-truncated, the values of $\theta$ are less informative, since they do not equal the sample mean, and hence have not been shown here.}
    \label{tab:xbar rho}
\end{table}
As expected, the mean number of transmissions is greater for males, regardless of cavitation status, and for infectors with cavitated TB, regardless of gender. Furthermore, male infectors with cavitated TB have the greatest number of transmissions on average, whilst female infectors with non-cavitated TB have the fewest. However, there is evidence that super-spreaders can emerge from any combination of gender and cavitation status, as $\rho > 1$ for all stratifications. Furthermore, the average values of $t_{20}$ were between 64\% and 80\% for all stratifications (see Table \ref{tab:t20}).\par
\begin{table}
    \centering
    \begin{tabular}{|l|l|l|l|}
         \hline
          & \textbf{Male} & \textbf{Female} & \textbf{Both}\\
         \hline
         \textbf{Cavitated} & 0.7934 & 0.7550 & 0.7968\\
          & (0.6131, 0.9293) & (0.4500, 0.9751) & (0.6283, 0.9357)\\
          \hline
         \textbf{Not cavitated} & 0.6794 & 0.6473 & 0.6794\\
          & (0.5016, 0.9015) & (0.4275, 0.8853) & (0.5335, 0.8617)\\
         \hline
         \textbf{Both} & 0.7795 & 0.7347 & 0.7751\\
          & (0.6040, 0.9159) & (0.4874, 0.9486) & (0.6298, 0.9010)\\
          \hline         
    \end{tabular}
    \caption{Mean and (minimum, maximum) values of $t_{20}$ by stratification across the 300 simulations.}
    \label{tab:t20}
\end{table}
Having said this, the average values for $t_{20}$ were greater for males than for females, regardless of whether cavitation status was stratified for, and greater for infectors with cavitated TB than those without cavitated TB, regardless of whether gender was stratified for. So, there is some weak evidence that super-spreaders are more likely to be male and have cavitated TB, but it is clear that super-spreaders can also be female and have non-cavitated TB.
\subsection{Counterfactual scenarios}
To determine to what extent certain differences between genders led to a higher TB burden in males, we compared the calibrated results to four counterfactual scenarios. In Scenario 1, we set all gender-specific parameter values to be equal for both genders (that is, we took an average of the male and female parameter values) and removed polygamy. In Scenario 2, we set any parameters involved in generating diagnosis delays equal for both genders (by averaging the mean values and inverse cumulative distribution functions for male and female diagnosis delays) but left other parameters unchanged. In Scenario 3, we set any parameters involved in determining the within-host disease progression equal for both genders (by averaging the male and female within-host parameter values) but left other parameters unchanged. In Scenario 4, we removed polygamous marriages but left other parameters unchanged. A summary of which parameters were set equal (or set to zero in the case of the proportion of polygamous marriages) for each counterfactual can be found in Table \ref{tab:counterfactual params}. A summary of the average total case numbers and male-to-female case ratios at the last time point of each simulation for the baseline scenario and each counterfactual scenario is given in Table \ref{tab:counterfactuals}.\par
\begin{figure}
    \centering
    \begin{subfigure}{0.45\linewidth}     
        \centering
        \includegraphics[width=\linewidth]{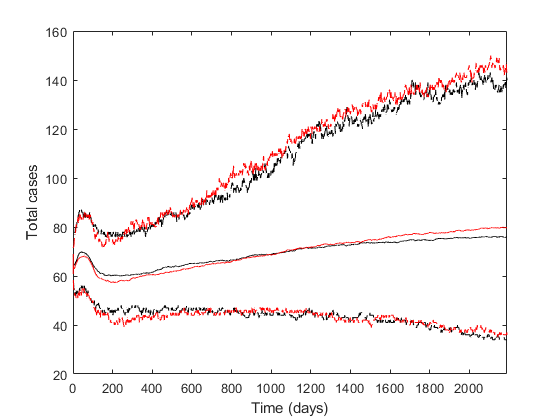}
        \label{fig:total all params}
        \caption{Scenario 1.}
    \end{subfigure}    
    \hfill
    \begin{subfigure}{0.45\linewidth}
        \centering
        \includegraphics[width=\linewidth]{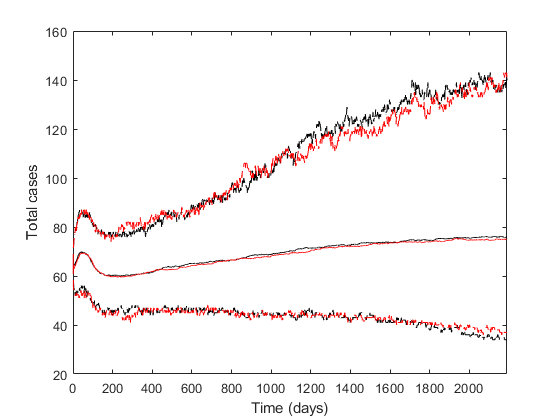}
        \label{fig:total diag delay}
        \caption{Scenario 2.}
    \end{subfigure}
    \begin{subfigure}{0.45\linewidth}
        \centering
        \includegraphics[width=\linewidth]{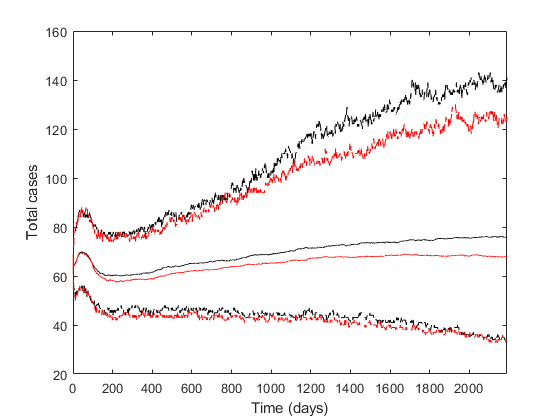}
        \label{fig:total within host}
        \caption{Scenario 3.}
    \end{subfigure}
    \hfill
    \begin{subfigure}{0.45\linewidth}
        \centering
        \includegraphics[width=\linewidth]{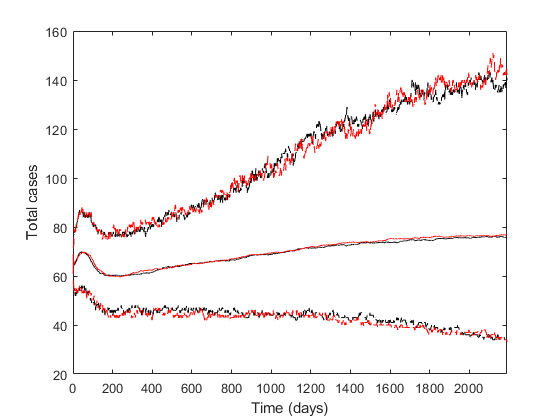}
        \label{fig:total polygamy}
        \caption{Scenario 4.}
    \end{subfigure}
    \caption{Summary plots of the total number of cases over 300 simulations of each counterfactual scenario, compared against the baseline calibrated scenario (i.e. the scenario with all gender differences included, presented in Section \ref{sec:calibrated results}). The solid lines indicate the average outcomes, and the dashed lines indicate the 95\% confidence intervals. The black lines indicate the baseline calibrated scenario, and the red lines indicate the counterfactual scenario under investigation.}
    \label{fig:counterfactuals total}
\end{figure}
\begin{figure}
    \centering
    \begin{subfigure}{0.45\linewidth}
        \centering
        \includegraphics[width=\linewidth]{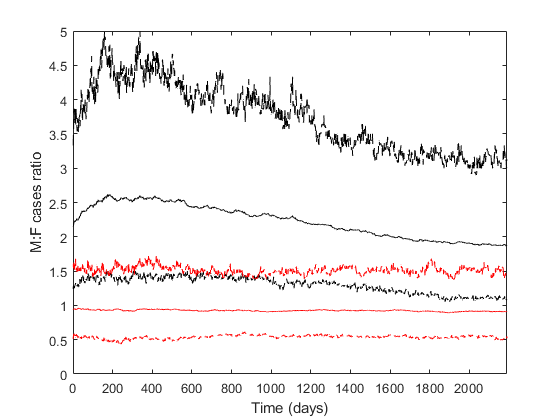}
        \label{fig:ratio all params}
        \caption{Scenario 1.}
    \end{subfigure}
    \hfill
    \begin{subfigure}{0.45\linewidth}
        \centering
        \includegraphics[width=\linewidth]{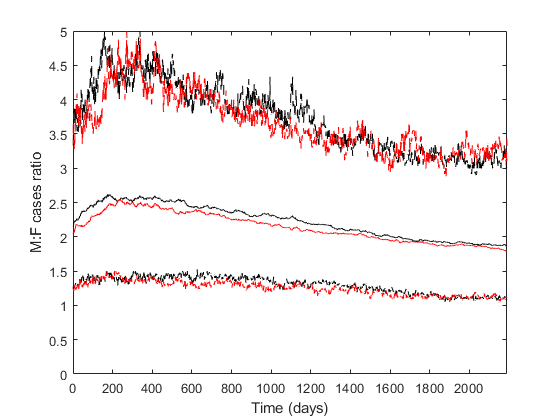}
        \label{fig:ratio diag delay}
        \caption{Scenario 2.}
    \end{subfigure}
    \begin{subfigure}{0.45\linewidth}
        \centering
        \includegraphics[width=\linewidth]{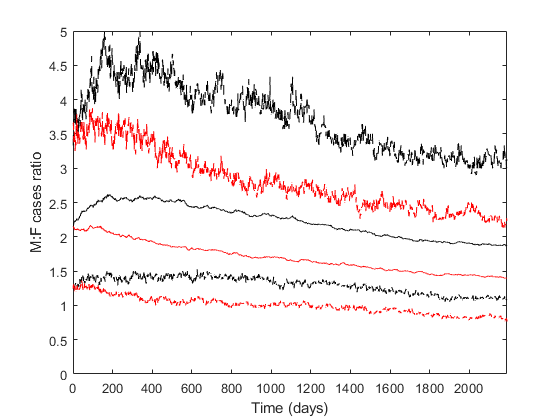}
        \label{fig:ratio within host}
        \caption{Scenario 3.}
    \end{subfigure}
    \hfill
    \begin{subfigure}{0.45\linewidth}
        \centering
        \includegraphics[width=\linewidth]{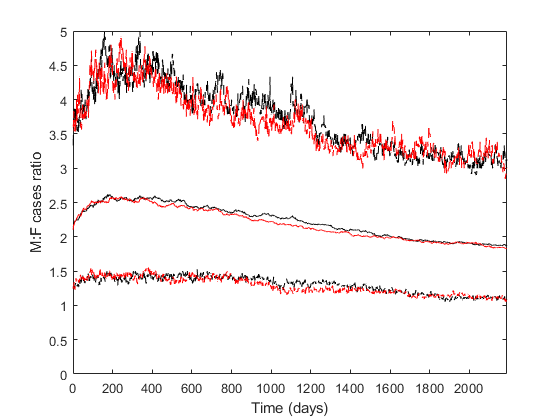}
        \label{fig:ratio polygamy}
        \caption{Scenario 4.}
    \end{subfigure}
    \caption{Summary plots of the male-to-female case ratio over 300 simulations of each counterfactual scenario, compared against the baseline calibrated scenario. The solid lines indicate the average outcomes, and the dashed lines indicate the 95\% confidence intervals. The black lines indicate the baseline calibrated scenario, and the red lines indicate the counterfactual scenario under investigation.}
    \label{fig:counterfactuals ratio}
\end{figure}

\begin{table}
    \centering
    \begin{tabular}{|l|l|l|l|l|}
    \hline
    \textbf{Parameter} & \textbf{1} & \textbf{2} & \textbf{3} & \textbf{4}\\
    \hline
    Male mean diagnosis delay & \checkmark & \checkmark & & \\
    Female mean diagnosis delay & \checkmark & \checkmark & & \\
    Male diagnosis delay ICDF & \checkmark & \checkmark & & \\ 
    Female diagnosis delay ICDF & \checkmark & \checkmark & & \\
    Male TB cases cavitated \% & \checkmark & & \checkmark & \\
    Female TB cases cavitated \% & \checkmark & & \checkmark & \\
    $\mathbb{P}$(male latent TB activation) & \checkmark & & \checkmark & \\
    $\mathbb{P}$(female latent TB activation) & \checkmark & & \checkmark & \\
    $\mathbb{P}$(male death via TB) & \checkmark & & \checkmark & \\
    $\mathbb{P}$(female death via TB) & \checkmark & & \checkmark & \\
    Polygamy proportion & \checkmark & & & \checkmark\\
    Males \% by place type & \checkmark & & & \\
    Females \% by place type & \checkmark & & & \\
    Household partner age gap & \checkmark & & & \\
    Case proportions by gender & \checkmark & & & \\
    WAIFW matrix by gender & \checkmark & & & \\
    \hline
    \end{tabular}
    \caption{List of the parameters set equal (or set to zero in the case of the number of polygamous marriages) for each counterfactual scenario. Scenario 1 is all gender-specific parameters set equal and polygamy removed; scenario 2 is diagnosis delays set equal; scenario 3 is within-host disease progression parameters set equal; scenario 4 is polygamy removed only. Abbreviations: ICDF, inverse cumulative distribution function; WAIFW, Who Acquired Infection From Whom.}
    \label{tab:counterfactual params}
\end{table}
As can be seen from the plots in Figures \ref{fig:counterfactuals total} and \ref{fig:counterfactuals ratio}, some counterfactual scenarios led to substantial differences in the total number of cases and the male-to-female case ratio, whilst others were broadly similar to the calibrated results. As expected, in Scenario 1, the male-to-female case ratio closely follows the male-to-female proportion across the whole population, which we had left as the calibrated parameter value. This led to a slight increase in the average number of cases: as the majority of cases were now in females, and we increased their diagnosis delay and probability of becoming sick when averaging these parameter values across both genders, their increase in cases was greater than the decrease in male cases that came from shorter male diagnosis delays and the lower probability of males becoming sick.\par
In Scenario 2, there was a slight decrease in total case numbers and the male-to-female case ratio. The average male diagnosis delay was reduced by approximately 2.5 days, less than 10\% of the previous value, so did not lead to a significant reduction in the number of transmissions before starting treatment; this would have also been somewhat offset by the slight increase in the average number of transmissions from females with active TB before they were diagnosed.\par
Scenario 3 saw the greatest reduction in the average total number of cases by the end of each simulation, and the male-to-female case ratio moved closer to parity. So men being given the same parameter values as women (i.e. being diagnosed more quickly on average, being less likely to have cavitated TB, and being less likely to develop active TB in the first place) appears to have reduced the impact of the greater probability of transmission from males with active TB relative to females with active TB. Initially, it may seem counter-intuitive that Scenario 3 sees a greater change from the baseline scenario than Scenario 1, despite the fact that the parameters altered in Scenario 3 are a subset of the parameters changed in Scenario 1. This would imply that the additional parameters in Scenario 1 that were different in value to those used in the baseline scenario counteract the differences in the parameters changed in Scenario 3. We believe that this was the case: the fact that women (who made up a larger proportion of the population) had longer diagnosis delays, increased probabilities of being assigned workplaces/schools (thereby increasing transmission possibilities), increased probabilities of cavitation and more probability of an infectious contact with either gender in Scenario 1 than in the baseline scenario, combined with a larger proportion of initial cases, countered the reduction in the probability of transmission from males, present in both Scenarios 1 and 3.\par
Only a small proportion of households would have had polygamous marriages, so the small changes in total case numbers and male-to-female case ratio for Scenario 4 seem intuitive: removing this would likely only reduce household transmissions in what would have otherwise been polygamous households.\par
In summary, all counterfactual scenarios appear to lead to a reduced male-to-female case ratio, compared to the baseline calibrated results. Scenarios 1 (all gender-specific parameters set equal) and 4 (polygamy removed) lead to an increase in the total number of cases compared to the baseline, while Scenarios 2 (diagnosis delays set equal) and 3 (within-host parameters set equal) lead to a decrease in total cases compared to the baseline.
\begin{table}
    \centering
    \begin{tabular}{|l|l|l|}
        \hline
         \textbf{Scenario} & \textbf{Total cases} & \textbf{M:F case ratio}\\
         \hline
         Baseline & $76.09 \pm 28.68$ & $1.88 \pm 0.55$\\
         All gender-specific parameters set equal (1) & $79.78 \pm 31.37$ & $0.91 \pm 0.23$\\
         Diagnosis delays set equal (2) & $75.00 \pm 27.87$ & $1.79 \pm 0.54$\\
         Within-host parameters set equal (3) & $68.11 \pm 23.44$ & $1.40 \pm 0.37$\\
         Polygamy removed (4) & $76.81 \pm 29.34$ & $1.83 \pm 0.57$\\
         \hline
    \end{tabular}
    \caption{A comparison of the mean and standard deviation of the total number of cases, and the mean and standard deviation of the male-to-female case ratio at the end of each simulation between the baseline scenario and each counterfactual scenario considered (with scenario numbers given in brackets in the Scenario column). All values are rounded to 2 decimal places.}
    \label{tab:counterfactuals}
\end{table}
\section{Discussion}
We have been able to successfully replicate the transmission dynamics of the ongoing TB endemic in Kampala by adapting the \texttt{Epiabm} framework first introduced by \citeauthor{Gallagher2024} \cite{Gallagher2024}. By altering the disease states to be more specific to TB, by evolving the population's socio-demography through ageing and reassignment to new households, workplaces and schools over time, and by stratifying the population by gender, we have generated case numbers and male-to-female TB case ratios that closely align with reality. Furthermore, we have found our model is capable of reproducing super-spreader events typically associated with TB transmission, as found in multiple previous studies (e.g. \cite{Ypma2013}, \cite{McCreesh2018}, \cite{Melsew2019}, \cite{Lee2020}, \cite{Teicher2023}, and \cite{Smith2023}). Finally, by considering four counterfactual scenarios, we provided evidence for and against some of the possible factors that may lead to the higher TB case burden in males. In this way, our model can support Uganda's National TB and Leprosy Program, and the Kampala Capital City Authority TB control strategies.\par
Our observation that super-spreaders can belong to either gender, and have both cavitary and non-cavitary TB, emphasises the importance of rapid diagnosis for anyone with active TB, regardless of their individual characteristics. Having said this, the fact that super-spreaders are more likely to be male and are more likely to have cavitary TB suggests the diagnosis of individuals with either, or both, of these characteristics should be prioritised. The counterfactual scenario where diagnosis delay was equalised (i.e. Scenario 2) did not lead to a significant reduction in the total number of cases, but we set the length of diagnosis delay equal to the average across both genders; it would be interesting to consider best-case counterfactual scenarios, such as no diagnosis delays for any individuals with active TB, and how large a reduction in case numbers could be achieved as a result, as future work. The counterfactual scenarios also highlight the issue of sex assortativity in adult contacts: the total number of cases and male-to-female case ratio were reduced when the within-host disease progression parameters were set equal, but the far greater likelihood of male-to-male transmission, compared to any other infector-infectee gender combination, meant the ratio did not approach parity.\par
Our model has some limitations. Perhaps the primary amongst these is that the agent-based model is computationally expensive. Consequently, it takes multiple hours to complete a single simulation when using a population the size of Kampala. As a result, we have had to reduce the size of the population and the period of time under investigation and we have had to use a supercomputer to complete all our simulations in a reasonable amount of time. Specifically, using the University of Bath's Nimbus cloud supercomputer, each simulation of a six-year time period typically took around 50 minutes to complete with the scaled-down population. These simplifying assumptions may cause additional complications: our results may only be applicable to populations of a similar size over time frames of a similar duration. Another limitation is that a number of parameter values have had to be estimated given the difficulty of observing the true values in the real-world data (the initial numbers of individuals in each compartment, for example). The counterfactual scenarios demonstrate the impact that altering certain key parameters can have on the results, but due to the complexity of the model, we have been unable to test the sensitivity of all model parameters, some of which may be having a substantial impact and could have feasibly taken alternative values. Additionally, as with other stochastic agent-based modelling frameworks, one needs to average a large number of simulations in order to be confident about the results; we have used the methodology proposed by \citeauthor{Hamis2021} to determine how many simulations are needed \cite{Hamis2021}, but a deterministic model would only require simulating once. Having said this, it is worth noting that deterministic modelling frameworks do not account for uncertainty in their results when provided with a given set of input variables and parameter values, something which our agent-based model is able to capture.
\label{sec:discussion}
\subsection{Future work}
With additional high-quality data, we could have added further features to our model. For example, we have assumed men in polygamous relationships have two wives, but if we had access to data that suggested otherwise (e.g. a probability distribution of the number of wives for Ugandan males), we could update the model accordingly. Having said this, Scenario 4 suggested polygamy has little impact on TB case numbers or the male-to-female case ratio, so it seems unlikely that increasing the number of wives in polygamous marriages in our model would change the lack of impact polygamy has on the output. Other features that could be implemented in future iterations of the model include immigration and emigration into and out of the city being modelled - \texttt{Epiabm} already has methods in place to deal with travellers in this regard - and different types of workplaces and spatial locations: there has been evidence that TB is more likely to be transmitted in certain places than others. For example, it has been shown that, in Dar es Salaam in Tanzania, TB is more likely to be spread in prisons and on public transport than in schools \cite{Hella2017}, so it would be interesting to see if this holds in other countries as well, especially as places with higher risk (e.g. prisons, public transport) are likely to have predominantly male workforces in some parts of the world. Although we have used a model structure with progression from latent TB to active TB closely matching Structure E from the classification proposed by \citeauthor{Menzies2018} \cite{Menzies2018} (with the structures presented in more detail in Appendix \ref{appendix: latent to active TB}), better contact tracing data could potentially allow us to identify which of these structures best fits the data and to alter our model accordingly. The counterfactual scenarios offer some indication of how sensitive the model output is to certain parameters, but a more thorough sensitivity analysis would reveal which parameters would be worth investigating further. Finally, we could consider additional counterfactuals that focus on some of the between-host factors we did not investigate in this study, such as equalising proportions of males and females in schools/workplaces only.
\subsection{Conclusion}
In this paper, we have demonstrated the impact gender has on TB transmission in Kampala. Specifically, through altering the \texttt{Epiabm} agent-based modelling framework, and using it to investigate the TB endemic over a six-year period, we have found that super-spreaders from both genders are likely to be the cause for the majority of transmissions, and that within-host factors, such as the greater likelihood of cavitary TB and of progressing from latent to active TB, appear to be the biggest factors in the high male TB burden. In addition, the high sex assortativity of contacts also appears to lead to the male-to-female case ratio remaining above parity, even when the within-host disease progression parameters are set equal. Our work therefore suggests both within-host and between-host factors influence the gender disparity in TB case numbers. Consequently, public health interventions looking to not only prevent cavitation, and the risk factors associated with it, but also to increase awareness of TB in men and the need for quick diagnosis and a reduction in social contacts whilst infectious (e.g. by self-isolating) are likely to prove impactful in reducing the severity of the TB epidemic in Kampala. Model outputs can inform targeted interventions such as increased male-focused screening and reducing diagnosis delays in men.
\section*{Acknowledgements}
JD thanks Ioana Bouros for their help in navigating the \texttt{Epiabm} model. This work makes use of the Nimbus cloud computer, and the authors gratefully acknowledge the University of Bath's Research Computing Group (\url{https://doi.org/10.15125/b6cd-s854}) for their support in this work.
\section*{Funding}
This work was supported by the Medical Research Council, United Kingdom [grant number MR/Y010124/1]. K.G. acknowledges funding from the EPSRC CDT in Sustainable Approaches to Biomedical Science: Responsible and Reproducible Research - SABS:R3 (EP/S024093/1).
\section*{CRediT authorship contribution statement}
\begin{itemize}
    \item \textbf{James W. G. Doran:} Conceptualization, Formal analysis, Investigation, Software, Validation, Visualization, Writing - original draft, Writing - review \& editing.
    \item \textbf{Dennis Mujuni:} Data curation, Writing - review \& editing.
    \item \textbf{Kit Gallagher:} Software, Writing - review \& editing.
    \item \textbf{Christian A. Yates:} Conceptualization, Supervision, Writing - review \& editing.
    \item \textbf{Ruth Bowness:} Conceptualization, Funding acquisition, Supervision, Writing - review \& editing.
\end{itemize}
\section*{Declaration of competing interest}
The authors declare the following financial interests/personal relationships which may be considered as potential competing interests: Ruth Bowness reports a relationship with Medical Research Council that includes: funding grants (grant number MR/Y010124/1).
\section*{Data statement}
The model code can be found at \url{https://github.com/jwgd93/EpiabmTB}. The data presented in this article can be found at the following links: \url{https://doi.org/10.5281/zenodo.18098364} for the consistency analysis data; \url{https://doi.org/10.5281/zenodo.18098766} for the calibrated results data; \url{https://doi.org/10.5281/zenodo.18098747} for the counterfactual scenarios data. These links will become live after publication.
\raggedright \printbibliography

@Article{W.H.O.2024,
  author = {{World Health Organization}},
  title  = {Global tuberculosis report 2024},
  year   = {2024},
  url    = {https://iris.who.int/bitstream/handle/10665/379339/9789240101531-eng.pdf?sequence=1},
}

@Article{Houben2016,
  author    = {Houben, Rein M. G. J. and Dodd, Peter J.},
  journal   = {PLOS Medicine},
  title     = {The global burden of latent tuberculosis infection: a re-estimation using mathematical modelling},
  year      = {2016},
  number    = {10},
  volume    = {13},
  publisher = {Public Library of Science San Francisco, CA USA},
}

@Article{Zwick2021,
  author    = {Zwick, Erin D. and Pepperell, Caitlin S. and Alagoz, Oguzhan},
  journal   = {Medical Decision Making},
  title     = {Representing tuberculosis transmission with complex contagion: an agent-based simulation modeling approach},
  year      = {2021},
  number    = {6},
  volume    = {41},
  publisher = {SAGE Publications Sage CA: Los Angeles, CA},
}

@Article{Prats2016,
  author    = {Prats, Clara and Montañola-Sales, Cristina and Gilabert-Navarro, Joan F. and Valls, Joaquim and Casanovas-Garcia, Josep and Vilaplana, Cristina and Cardona, Pere-Joan and López, Daniel},
  journal   = {Frontiers in Microbiology},
  title     = {Individual-based modeling of tuberculosis in a user-friendly interface: understanding the epidemiological role of population heterogeneity in a city},
  year      = {2016},
  volume    = {6},
  publisher = {Frontiers},
}

@Article{Kasaie2015,
  author    = {Kasaie, Parastu and Mathema, Barun and Kelton, W. David and Azman, Andrew S. and Pennington, Jeff and Dowdy, David W.},
  journal   = {PLOS One},
  title     = {A novel tool improves existing estimates of recent tuberculosis transmission in settings of sparse data collection},
  year      = {2015},
  number    = {12},
  volume    = {10},
  publisher = {Public Library of Science San Francisco, CA USA},
}

@Article{Shrestha2017,
  author    = {Shrestha, Sourya and Hill, Andrew N. and Marks, Suzanne M. and Dowdy, David W.},
  journal   = {American Journal of Respiratory and Critical Care Medicine},
  title     = {Comparing drivers and dynamics of tuberculosis in California, Florida, New York, and Texas},
  year      = {2017},
  number    = {8},
  volume    = {196},
  publisher = {American Thoracic Society},
}

@Article{Guzzetta2011,
  author    = {Guzzetta, Giorgio and Ajelli, Marco and Yang, Zhenhua and Merler, Stefano and Furlanello, Cesare and Kirschner, Denise},
  journal   = {Journal of Theoretical Biology},
  title     = {Modeling socio-demography to capture tuberculosis transmission dynamics in a low burden setting},
  year      = {2011},
  volume    = {289},
  publisher = {Elsevier},
}

@InProceedings{Udall2023,
  author    = {Udall, Chrystal Bianca and Searle, Christa},
  booktitle = {Operations Research Forum},
  title     = {Modelling the Spread of Tuberculosis in South African Informal Settlements},
  year      = {2023},
  number    = {1},
  publisher = {Springer},
  volume    = {4},
}

@Article{Tian2013,
  author    = {Tian, Yuan and Osgood, Nathaniel D. and Al-Azem, Assaad and Hoeppner, Vernon H.},
  journal   = {Health Education \& Behavior},
  title     = {Evaluating the effectiveness of contact tracing on tuberculosis outcomes in Saskatchewan using individual-based modeling},
  year      = {2013},
  number    = {1_suppl},
  volume    = {40},
  publisher = {SAGE Publications Sage CA: Los Angeles, CA},
}

@Book{Ferguson2020,
  author    = {Ferguson, Neil M. and Laydon, Daniel and Nedjati-Gilani, Gemma and Imai, Natsuko and Ainslie, Kylie and Baguelin, Marc and Bhatia, Sangeeta and Boonyasiri, Adhiratha and Cucunubá, Zulma and Cuomo-Dannenburg, Gina},
  publisher = {Imperial College London London},
  title     = {Report 9: Impact of non-pharmaceutical interventions (NPIs) to reduce COVID19 mortality and healthcare demand},
  year      = {2020},
  volume    = {16},
}

@Article{Ragonnet2019,
  author    = {Ragonnet, Romain and Trauer, James M. and Geard, Nicholas and Scott, Nick and McBryde, Emma S.},
  journal   = {BMC Medicine},
  title     = {Profiling Mycobacterium tuberculosis transmission and the resulting disease burden in the five highest tuberculosis burden countries},
  year      = {2019},
  volume    = {17},
  publisher = {Springer},
}

@Article{DeEspindola2011,
  author    = {De Espíndola, Aquino L. and Bauch, Chris T. and Cabella, Brenno C. Troca and Martinez, Alexandre Souto},
  journal   = {Journal of Statistical Mechanics: Theory and Experiment},
  title     = {An agent-based computational model of the spread of tuberculosis},
  year      = {2011},
  number    = {05},
  volume    = {2011},
  publisher = {IOP Publishing},
}

@Article{Kubjane2023,
  author    = {Kubjane, Mmamapudi and Cornell, Morna and Osman, Muhammad and Boulle, Andrew and Johnson, Leigh F.},
  journal   = {Scientific Reports},
  title     = {Drivers of sex differences in the South African adult tuberculosis incidence and mortality trends, 1990–2019},
  year      = {2023},
  number    = {1},
  volume    = {13},
  publisher = {Nature Publishing Group UK London},
}

@Article{Gallagher2024,
  author  = {Gallagher, Kit and Bouros, Ioana and Fan, Nicholas and Hayman, Elizabeth and Heirene, Luke and Lamirande, Patricia and Lemenuel-Diot, Annabelle and Lambert, Ben and Gavaghan, David and Creswell, Richard},
  journal = {Journal of Open Research Software},
  title   = {Epidemiological agent-based modelling software (epiabm)},
  year    = {2024},
  number  = {1},
  volume  = {12},
}

@Article{Ferguson2006,
  author    = {Ferguson, Neil M. and Cummings, Derek A. T. and Fraser, Christophe and Cajka, James C. and Cooley, Philip C. and Burke, Donald S.},
  journal   = {Nature},
  title     = {Strategies for mitigating an influenza pandemic},
  year      = {2006},
  number    = {7101},
  volume    = {442},
  publisher = {Nature Publishing Group},
}

@Article{Ferguson2005,
  author    = {Ferguson, Neil M. and Cummings, Derek A. T. and Cauchemez, Simon and Fraser, Christophe and Riley, Steven and Meeyai, Aronrag and Iamsirithaworn, Sopon and Burke, Donald S.},
  journal   = {Nature},
  title     = {Strategies for containing an emerging influenza pandemic in Southeast Asia},
  year      = {2005},
  number    = {7056},
  volume    = {437},
  publisher = {Nature Publishing Group},
}

@Article{Halloran2008,
  author    = {Halloran, M. Elizabeth and Ferguson, Neil M. and Eubank, Stephen and Longini, Ira M. and Cummings, Derek A. T. and Lewis, Bryan and Xu, Shufu and Fraser, Christophe and Vullikanti, Anil and Germann, Timothy C.},
  journal   = {Proceedings of the National Academy of Sciences},
  title     = {Modeling targeted layered containment of an influenza pandemic in the United States},
  year      = {2008},
  number    = {12},
  volume    = {105},
  publisher = {National Acad Sciences},
}

@Article{Samuels1965,
  author    = {Samuels, J.O.H.N. M.},
  journal   = {The Review of Economic Studies},
  title     = {Size and the growth of firms},
  year      = {1965},
  number    = {2},
  volume    = {32},
  publisher = {Wiley-Blackwell},
}

@Article{Goyette2014,
  author    = {Goyette, Jonathan},
  journal   = {The European Journal of Development Research},
  title     = {The determinants of the size distribution of firms in Uganda},
  year      = {2014},
  volume    = {26},
  publisher = {Springer},
}

@Article{Nhamoyebonde2014,
  author    = {Nhamoyebonde, Shepherd and Leslie, Alasdair},
  journal   = {The Journal of Infectious Diseases},
  title     = {Biological differences between the sexes and susceptibility to tuberculosis},
  year      = {2014},
  publisher = {JSTOR},
}

@Article{Blower1995,
  author    = {Blower, Sally M. and Mclean, Angela R. and Porco, Travis C. and Small, Peter M. and Hopewell, Philip C. and Sanchez, Melissa A. and Moss, Andrew R.},
  journal   = {Nature Medicine},
  title     = {The intrinsic transmission dynamics of tuberculosis epidemics},
  year      = {1995},
  number    = {8},
  volume    = {1},
  publisher = {Nature Publishing Group},
}

@Article{Jarosz2021,
  author    = {Jarosz, Beth},
  journal   = {Population Research and Policy Review},
  title     = {Poisson distribution: A model for estimating households by household size},
  year      = {2021},
  number    = {2},
  volume    = {40},
  publisher = {Springer},
}

@Article{Wallinga2006,
  author    = {Wallinga, Jacco and Teunis, Peter and Kretzschmar, Mirjam},
  journal   = {American Journal of Epidemiology},
  title     = {Using data on social contacts to estimate age-specific transmission parameters for respiratory-spread infectious agents},
  year      = {2006},
  number    = {10},
  volume    = {164},
  publisher = {Oxford University Press},
}

@Article{Borgdorff2011,
  author    = {Borgdorff, Martien W. and Sebek, Maruschka and Geskus, Ronald B. and Kremer, Kristin and Kalisvaart, Nico and van Soolingen, Dick},
  journal   = {International Journal of Epidemiology},
  title     = {The incubation period distribution of tuberculosis estimated with a molecular epidemiological approach},
  year      = {2011},
  number    = {4},
  volume    = {40},
  publisher = {Oxford University Press},
}

@Article{Tiemersma2011,
  author    = {Tiemersma, Edine W. and van der Werf, Marieke J. and Borgdorff, Martien W. and Williams, Brian G. and Nagelkerke, Nico J. D.},
  journal   = {PLOS One},
  title     = {Natural history of tuberculosis: duration and fatality of untreated pulmonary tuberculosis in HIV negative patients: a systematic review},
  year      = {2011},
  number    = {4},
  volume    = {6},
  publisher = {Public Library of Science San Francisco, USA},
}

@Article{UBOS2021,
  author = {{Uganda Bureau of Statistics}},
  title  = {Uganda National Household Survey 2019/2020},
  year   = {2021},
}

@Misc{UN2024,
  author = {{United Nations}},
  note   = {Processed by Our World in Data},
  title  = {World Population Prospects},
  year   = {2024},
  url    = {https://ourworldindata.org/grapher/population-by-five-year-age-group?country=~UGA},
}

@Misc{MES2017,
  author = {{Ministry of Education \& Sports}},
  title  = {Education Abstract 2017},
  year   = {2017},
}

@Article{Shaikh2021,
  author    = {Shaikh, Ambreen and Sriraman, Kalpana and Vaswani, Smriti and Oswal, Vikas and Rao, Sudha and Mistry, Nerges},
  journal   = {Scientific Reports},
  title     = {Early phase of effective treatment induces distinct transcriptional changes in Mycobacterium tuberculosis expelled by pulmonary tuberculosis patients},
  year      = {2021},
  number    = {1},
  volume    = {11},
  publisher = {Nature Publishing Group UK London},
}

@Article{Wang2024,
  author    = {Wang, Si and Cao, Hui},
  journal   = {Journal of Biological Dynamics},
  title     = {The dynamics of tuberculosis transmission model with different genders},
  year      = {2024},
  number    = {1},
  volume    = {18},
  publisher = {Taylor & Francis},
}

@Article{KisselevskayaBabinina2018,
  author    = {Kisselevskaya-Babinina, Viktoriya Yaroslavovna and Sannikova, Tat'yana Evgen'evna and Romanyukha, Aleksei Alekseevich and Karkach, Arsenii Sergeevich},
  journal   = {Matematicheskaya Biologiya i Bioinformatika},
  title     = {Modeling of gender differences in tuberculosis prevalence},
  year      = {2018},
  number    = {2},
  volume    = {13},
  publisher = {Institute of Mathematical Problems of Biology, Russian Academy of Sciences},
}

@Article{Obeagu2023,
  author    = {Obeagu, Emmanuel I.},
  journal   = {Health Science Reports},
  title     = {Tuberculosis diagnostic and treatment delays among patients in Uganda},
  year      = {2023},
  number    = {11},
  volume    = {6},
  publisher = {Wiley Online Library},
}

@Article{Kakumba2023,
  author = {Kakumba, Makanga Ronald},
  title  = {Uganda a continental extreme in rejection of people in same-sex relationships},
  year   = {2023},
}

@Article{VanDenDriessche2007,
  author    = {Van Den Driessche, P. and Wang, Lin and Zou, Xingfu},
  journal   = {Mathematical Biosciences \& Engineering},
  title     = {Modeling diseases with latency and relapse},
  year      = {2007},
  number    = {2},
  volume    = {4},
  publisher = {Mathematical Biosciences & Engineering},
}

@Misc{UBOS2018,
  author = {{Uganda Bureau of Statistics}},
  title  = {Uganda Demographic and Health Survey 2016},
  year   = {2018},
}

@Misc{UIS2023,
  author = {{UNESCO Institute for Statistics}},
  title  = {Official entrance age and theoretical duration by level of education (years)},
  year   = {2023},
}

@Article{Kelly2003,
  author    = {Kelly, Robert J. and Gray, Ronald H. and Sewankambo, Nelson K. and Serwadda, David and Wabwire-Mangen, Fred and Lutalo, Tom and Wawer, Maria J.},
  journal   = {JAIDS Journal of Acquired Immune Deficiency Syndromes},
  title     = {Age differences in sexual partners and risk of HIV-1 infection in rural Uganda},
  year      = {2003},
  number    = {4},
  volume    = {32},
  publisher = {LWW},
}

@Article{Fox2016,
  author    = {Fox, Gregory J. and Orlova, Marianna and Schurr, Erwin},
  journal   = {PLOS Pathogens},
  title     = {Tuberculosis in newborns: the lessons of the “Lübeck Disaster”(1929–1933)},
  year      = {2016},
  number    = {1},
  volume    = {12},
  publisher = {Public Library of Science San Francisco, CA USA},
}

@Article{JimenezCorona2006,
  author    = {Jimenez-Corona, Maria-Eugenia and Garcia-Garcia, Lourdes and DeRiemer, Kathryn and Ferreyra-Reyes, Leticia and Bobadilla-del-Valle, Miriam and Cano-Arellano, Bulmaro and Canizales-Quintero, Sergio and Martinez-Gamboa, Areli and Small, P. M. and Sifuentes-Osornio, Jose},
  journal   = {Thorax},
  title     = {Gender differentials of pulmonary tuberculosis transmission and reactivation in an endemic area},
  year      = {2006},
  number    = {4},
  volume    = {61},
  publisher = {BMJ Publishing Group Ltd},
}

@Article{Gao2017,
  author    = {Gao, Lei and Li, Xiangwei and Liu, Jianmin and Wang, Xinhua and Lu, Wei and Bai, Liqiong and Xin, Henan and Zhang, Haoran and Li, Hengjing and Zhang, Zongde},
  journal   = {The Lancet Infectious Diseases},
  title     = {Incidence of active tuberculosis in individuals with latent tuberculosis infection in rural China: follow-up results of a population-based, multicentre, prospective cohort study},
  year      = {2017},
  number    = {10},
  volume    = {17},
  publisher = {Elsevier},
}

@Article{Urbanowski2020,
  author    = {Urbanowski, Michael E. and Ordonez, Alvaro A. and Ruiz-Bedoya, Camilo A. and Jain, Sanjay K. and Bishai, William R.},
  journal   = {The Lancet Infectious Diseases},
  title     = {Cavitary tuberculosis: the gateway of disease transmission},
  year      = {2020},
  number    = {6},
  volume    = {20},
  publisher = {Elsevier},
}

@Article{Balogun2021,
  author  = {Balogun, O. O. and Fawole, A. and Osemwinyen, E. and Balogun, B.},
  journal = {J Infect Dis Epidemiol},
  title   = {Predictors of Pulmonary Cavitation among Tuberculosis Patients},
  year    = {2021},
  volume  = {7},
}

@Article{Gadkowski2008,
  author    = {Gadkowski, L. Beth and Stout, Jason E.},
  journal   = {Clinical Microbiology Reviews},
  title     = {Cavitary pulmonary disease},
  year      = {2008},
  number    = {2},
  volume    = {21},
  publisher = {Am Soc Microbiol},
}

@Article{Zhang2016,
  author    = {Zhang, Liqun and Pang, Yu and Yu, Xia and Wang, Yufeng and Lu, Jie and Gao, Mengqiu and Huang, Hairong and Zhao, Yanlin},
  journal   = {Emerging Microbes \& Infections},
  title     = {Risk factors for pulmonary cavitation in tuberculosis patients from China},
  year      = {2016},
  number    = {1},
  volume    = {5},
  publisher = {Taylor & Francis},
}

@Misc{UNDESA2022,
  author = {{United Nations, Department of Economic and Social Affairs, Population Division}},
  title  = {Database on Household Size and Composition 2022},
  year   = {2022},
}

@Article{Mwanga2021,
  author  = {Mwanga, Mastullah Ashah and Paul, Shimiyu and Lajul, Wilfred},
  journal = {African Journal of Governance and Public Leadership},
  title   = {Governance challenges to women’s realisation of the right to sexual and reproductive health: the case of women in polygamous marriages in Uganda},
  year    = {2021},
  number  = {1},
  volume  = {1},
}

@Misc{UBOS2016,
  author = {{Uganda Bureau of Statistics}},
  title  = {The National Population and Housing Census 2014 – Main Report},
  year   = {2016},
}

@Article{Prem2017,
  author    = {Prem, Kiesha and Cook, Alex R. and Jit, Mark},
  journal   = {PLOS Computational Biology},
  title     = {Projecting social contact matrices in 152 countries using contact surveys and demographic data},
  year      = {2017},
  number    = {9},
  volume    = {13},
  publisher = {Public Library of Science San Francisco, CA USA},
}

@Misc{UIS2017,
  author = {{UNESCO Institute for Statistics}},
  title  = {Enrolment by level of education},
  year   = {2017},
}

@Misc{UIS2017a,
  author = {{UNESCO Institute for Statistics}},
  title  = {Number of teachers by teaching level of education},
  year   = {2017},
}

@Misc{UBOS2024,
  author = {{Uganda Bureau of Statistics}},
  title  = {The National Population and Housing Census 2024 – Main Report},
  year   = {2024},
}

@Misc{ILO2021,
  author = {{International Labour Organization}},
  title  = {Labour force participation rate by sex and age (\%) - Annual},
  year   = {2021},
}

@Article{Emery2021,
  author    = {Emery, Jon C. and Richards, Alexandra S. and Dale, Katie D. and McQuaid, C. Finn and White, Richard G. and Denholm, Justin T. and Houben, Rein M. G. J.},
  journal   = {Proceedings of the Royal Society B},
  title     = {Self-clearance of Mycobacterium tuberculosis infection: implications for lifetime risk and population at-risk of tuberculosis disease},
  year      = {2021},
  number    = {1943},
  volume    = {288},
  publisher = {The Royal Society},
}

@Misc{WHO2023,
  author = {{WHO African Region}},
  title  = {Tuberculosis in the WHO African Region: 2023 progress update},
  year   = {2023},
}

@Misc{WHO2024,
  author = {{World Health Organization}},
  title  = {Global Health Observatory},
  year   = {2024},
}

@Article{Colangeli2018,
  author    = {Colangeli, Roberto and Jedrey, Hannah and Kim, Soyeon and Connell, Roy and Ma, Shuyi and Chippada Venkata, Uma D. and Chakravorty, Soumitesh and Gupta, Aditi and Sizemore, Erin E. and Diem, Lois},
  journal   = {New England Journal of Medicine},
  title     = {Bacterial factors that predict relapse after tuberculosis therapy},
  year      = {2018},
  number    = {9},
  volume    = {379},
  publisher = {Mass Medical Soc},
}

@Article{Fromsa2024,
  author    = {Fromsa, Abebe and Willgert, Katriina and Srinivasan, Sreenidhi and Mekonnen, Getnet and Bedada, Wegene and Gumi, Balako and Lakew, Matios and Tadesse, Biniam and Bayissa, Berecha and Sirak, Asegedech},
  journal   = {Science},
  title     = {BCG vaccination reduces bovine tuberculosis transmission, improving prospects for elimination},
  year      = {2024},
  number    = {6690},
  volume    = {383},
  publisher = {American Association for the Advancement of Science},
}

@Article{Roy2014,
  author    = {Roy, Anjana and Eisenhut, Michael and Harris, R. J. and Rodrigues, L. C. and Sridhar, Saranya and Habermann, Stephanie and Snell, Luke and Mangtani, Punam and Adetifa, Ifedayo and Lalvani, Ajit},
  journal   = {BMJ},
  title     = {Effect of BCG vaccination against Mycobacterium tuberculosis infection in children: systematic review and meta-analysis},
  year      = {2014},
  volume    = {349},
  publisher = {British Medical Journal Publishing Group},
}

@Misc{WHOUNICEF2022,
  author = {{WHO/UNICEF}},
  title  = {Bacillus Calmette–Guérin (BCG) vaccination coverage},
  year   = {2022},
  url    = {https://immunizationdata.who.int/global/wiise-detail-page/bacillus-calmette-guérin-(bcg)-vaccination-coverage?CODE=UGA&YEAR=},
}

@Article{Buregyeya2014,
  author    = {Buregyeya, Esther and Criel, Bart and Nuwaha, Fred and Colebunders, Robert},
  journal   = {BMC Public Health},
  title     = {Delays in diagnosis and treatment of pulmonary tuberculosis in Wakiso and Mukono districts, Uganda},
  year      = {2014},
  volume    = {14},
  publisher = {Springer},
}

@Article{Hamis2021,
  author    = {Hamis, Sara and Stratiev, Stanislav and Powathil, Gibin G.},
  journal   = {THE PHYSICS OF CANCER: Research Advances},
  title     = {Uncertainty and sensitivity analyses methods for agent-based mathematical models: An introductory review},
  year      = {2021},
  publisher = {World Scientific},
}

@Article{Logitharajah2008,
  author    = {Logitharajah, Pavithra and Kampmann, Beate},
  journal   = {BMJ},
  title     = {Tuberculosis in children},
  year      = {2008},
  volume    = {337},
  publisher = {British Medical Journal Publishing Group},
}

@Article{Jjuuko2013,
  author    = {Jjuuko, Adrian},
  journal   = {Human Rights, Sexual Orientation and Gender Identity in The Commonwealth: Struggles for Decriminalisation and Change},
  title     = {The incremental approach: Uganda’s struggle for the decriminalisation of homosexuality},
  year      = {2013},
  publisher = {School of Advanced Study, University of London London},
}

@Article{CastilloChavez1997,
  author    = {Castillo-Chavez, Carlos and Feng, Zhilan},
  journal   = {Journal of Mathematical Biology},
  title     = {To treat or not to treat: the case of tuberculosis},
  year      = {1997},
  volume    = {35},
  publisher = {Springer},
}

@Article{Feng2001,
  author    = {Feng, Zhilan and Huang, Wenzhang and Castillo-Chavez, Carlos},
  journal   = {Journal of Dynamics and Differential Equations},
  title     = {On the role of variable latent periods in mathematical models for tuberculosis},
  year      = {2001},
  volume    = {13},
  publisher = {Springer},
}

@Article{Schwalb,
  author  = {Schwalb, Alvaro and Dodd, Pete and Rickman, Hannah M. and Ugarte-Gil, César and Horton, Katherine C. and Houben, Rein M. G. J.},
  journal = {Available at SSRN 5017943},
  title   = {Estimating the Global Burden of Viable Mycobacterium Tuberculosis Infection},
}

@Article{Whalen2011,
  author    = {Whalen, Christopher C. and Zalwango, Sarah and Chiunda, Allan and Malone, LaShaunda and Eisenach, Kathleen and Joloba, Moses and Boom, W. Henry and Mugerwa, Roy},
  journal   = {PLOS One},
  title     = {Secondary attack rate of tuberculosis in urban households in Kampala, Uganda},
  year      = {2011},
  number    = {2},
  volume    = {6},
  publisher = {Public Library of Science San Francisco, USA},
}

@Article{Bui2024,
  author    = {Bui, Viet Long and Hughes, Angus E. and Ragonnet, Romain and Meehan, Michael T. and Henderson, Alec and McBryde, Emma S. and Trauer, James M.},
  journal   = {BMC Infectious Diseases},
  title     = {Agent-based modelling of Mycobacterium tuberculosis transmission: a systematic review},
  year      = {2024},
  number    = {1},
  volume    = {24},
  publisher = {Springer},
}

@Article{Menzies2018,
  author    = {Menzies, Nicolas A. and Wolf, Emory and Connors, David and Bellerose, Meghan and Sbarra, Alyssa N. and Cohen, Ted and Hill, Andrew N. and Yaesoubi, Reza and Galer, Kara and White, Peter J.},
  journal   = {The Lancet Infectious Diseases},
  title     = {Progression from latent infection to active disease in dynamic tuberculosis transmission models: a systematic review of the validity of modelling assumptions},
  year      = {2018},
  number    = {8},
  volume    = {18},
  publisher = {Elsevier},
}

@Article{Aceng2024,
  author    = {Aceng, Freda Loy and Kabwama, Steven Ndugwa and Ario, Alex Riolexus and Etwom, Alfred and Turyahabwe, Stavia and Mugabe, Frank Rwabinumi},
  journal   = {BMC Infectious Diseases},
  title     = {Spatial distribution and temporal trends of tuberculosis case notifications, Uganda: a ten-year retrospective analysis (2013–2022)},
  year      = {2024},
  number    = {1},
  volume    = {24},
  publisher = {Springer},
}

@Misc{UN2024a,
  author = {{United Nations}},
  note   = {Processed by Our World in Data},
  title  = {World Population Prospects},
  year   = {2024},
  url    = {https://ourworldindata.org/explorers/global-health?country=~UGA&Health+Area=Deaths+and+DALYs&Indicator=Total+deaths&Metric=Rate&Source=UN+WPP},
}

@Article{Miller2021,
  author    = {Miller, Paige B. and Zalwango, Sarah and Galiwango, Ronald and Kakaire, Robert and Sekandi, Juliet and Steinbaum, Lauren and Drake, John M. and Whalen, Christopher C. and Kiwanuka, Noah},
  journal   = {BMC Infectious Diseases},
  title     = {Association between tuberculosis in men and social network structure in Kampala, Uganda},
  year      = {2021},
  volume    = {21},
  publisher = {Springer},
}

@Article{Koch1982,
  author    = {Koch, Robert},
  journal   = {Reviews of Infectious Diseases},
  title     = {The etiology of tuberculosis},
  year      = {1982},
  number    = {6},
  volume    = {4},
  publisher = {The University of Chicago Press},
}

@Article{Horton2016,
  author    = {Horton, Katherine C. and MacPherson, Peter and Houben, Rein M. G. J. and White, Richard G. and Corbett, Elizabeth L.},
  journal   = {PLOS Medicine},
  title     = {Sex differences in tuberculosis burden and notifications in low-and middle-income countries: a systematic review and meta-analysis},
  year      = {2016},
  number    = {9},
  volume    = {13},
  publisher = {Public Library of Science San Francisco, CA USA},
}

@Article{Horton2020,
  author  = {Horton, Katherine C. and Hoey, Anne L. and Béraud, Guillaume and Corbett, Elizabeth L. and White, Richard G.},
  journal = {Emerging Infectious Diseases},
  title   = {Systematic review and meta-analysis of sex differences in social contact patterns and implications for tuberculosis transmission and control},
  year    = {2020},
  number  = {5},
  volume  = {26},
}

@Article{Richards2025,
  author    = {Richards, Alexandra S. and Phiri, Mphatso D. and Nidoi, Jasper and Chakaya, Jeremiah and MacPherson, Peter and Kirenga, Bruce J. and Bimba, John S. and Ugwu, Chukwuebuka and Pola, Rhoda and Squire, S. Bertel},
  journal   = {medRxiv},
  title     = {Population level impact of increasing tuberculosis treatment coverage and addressing determinants of risk in men: a modelling study in Kenya, Malawi, Nigeria, and Uganda},
  year      = {2025},
  publisher = {Cold Spring Harbor Laboratory Press},
}

@Article{Goyal2022,
  author    = {Goyal, Ashish and Reeves, Daniel B. and Schiffer, Joshua T.},
  journal   = {Journal of the Royal Society Interface},
  title     = {Multi-scale modelling reveals that early super-spreader events are a likely contributor to novel variant predominance},
  year      = {2022},
  number    = {189},
  volume    = {19},
  publisher = {The Royal Society},
}

@Article{Melsew2019,
  author    = {Melsew, Yayehirad A. and Gambhir, Manoj and Cheng, Allen C. and McBryde, Emma S. and Denholm, Justin T. and Tay, Ee Laine and Trauer, James M.},
  journal   = {BMC Infectious Diseases},
  title     = {The role of super-spreading events in Mycobacterium tuberculosis transmission: evidence from contact tracing},
  year      = {2019},
  number    = {1},
  volume    = {19},
  publisher = {Springer},
}

@Article{Teicher2023,
  author    = {Teicher, Amir},
  journal   = {The Lancet Infectious Diseases},
  title     = {Super-spreaders: a historical review},
  year      = {2023},
  number    = {10},
  volume    = {23},
  publisher = {Elsevier},
}

@Article{Ypma2013,
  author    = {Ypma, Rolf J. F. and Altes, Hester Korthals and van Soolingen, Dick and Wallinga, Jacco and Van Ballegooijen, W. Marijn},
  journal   = {Epidemiology},
  title     = {A sign of superspreading in tuberculosis: highly skewed distribution of genotypic cluster sizes},
  year      = {2013},
  number    = {3},
  volume    = {24},
  publisher = {LWW},
}

@Article{Lee2020,
  author    = {Lee, Robyn S. and Proulx, Jean-François and McIntosh, Fiona and Behr, Marcel A. and Hanage, William P.},
  journal   = {Elife},
  title     = {Previously undetected super-spreading of Mycobacterium tuberculosis revealed by deep sequencing},
  year      = {2020},
  volume    = {9},
  publisher = {eLife Sciences Publications, Ltd},
}

@Article{Smith2023,
  author    = {Smith, Jonathan P. and Cohen, Ted and Dowdy, David and Shrestha, Sourya and Gandhi, Neel R. and Hill, Andrew N.},
  journal   = {American Journal of Epidemiology},
  title     = {Quantifying Mycobacterium tuberculosis transmission dynamics across global settings: a systematic analysis},
  year      = {2023},
  number    = {1},
  volume    = {192},
  publisher = {Oxford University Press},
}

@Article{McCreesh2018,
  author    = {McCreesh, Nicky and White, Richard G.},
  journal   = {Scientific Reports},
  title     = {An explanation for the low proportion of tuberculosis that results from transmission between household and known social contacts},
  year      = {2018},
  number    = {1},
  volume    = {8},
  publisher = {Nature Publishing Group UK London},
}

@Article{Vargha2000,
  author    = {Vargha, András and Delaney, Harold D.},
  journal   = {Journal of Educational and Behavioral Statistics},
  title     = {A critique and improvement of the CL common language effect size statistics of McGraw and Wong},
  year      = {2000},
  number    = {2},
  volume    = {25},
  publisher = {Sage Publications Sage CA: Los Angeles, CA},
}

@Article{Alden2013,
  author    = {Alden, Kieran and Read, Mark and Timmis, Jon and Andrews, Paul S. and Veiga-Fernandes, Henrique and Coles, Mark},
  journal   = {PLOS Computational Biology},
  title     = {Spartan: a comprehensive tool for understanding uncertainty in simulations of biological systems},
  year      = {2013},
  number    = {2},
  volume    = {9},
  publisher = {Public Library of Science San Francisco, USA},
}

@Article{Galvani2005,
  author    = {Galvani, Alison P. and May, Robert M.},
  journal   = {Nature},
  title     = {Dimensions of superspreading},
  year      = {2005},
  number    = {7066},
  volume    = {438},
  publisher = {Nature Publishing Group UK London},
}

@Misc{UBOS2019,
  author = {{Uganda Bureau of Statistics}},
  title  = {Table 1.1 : Estimated Land Area and Projected Population by Sex by Lower Local Government},
  year   = {2019},
}

@Article{Hunter2022,
  author    = {Hunter, Elizabeth and Kelleher, John D.},
  journal   = {Algorithms},
  title     = {Validating and testing an agent-based model for the spread of COVID-19 in Ireland},
  year      = {2022},
  number    = {8},
  volume    = {15},
  publisher = {MDPI},
}

@Article{Boum2014,
  author    = {Boum, Yap and Atwine, Daniel and Orikiriza, Patrick and Assimwe, Justus and Page, Anne-Laure and Mwanga-Amumpaire, Juliet and Bonnet, Maryline},
  journal   = {BMC Infectious Diseases},
  title     = {Male Gender is independently associated with pulmonary tuberculosis among sputum and non-sputum producers people with presumptive tuberculosis in Southwestern Uganda},
  year      = {2014},
  number    = {1},
  volume    = {14},
  publisher = {Springer},
}

@Misc{WHO2024a,
  author = {{World Health Organization}},
  note   = {With minor processing by Our World in Data},
  title  = {Global Tuberculosis Report - Burden Estimates},
  year   = {2024},
  url    = {https://ourworldindata.org/grapher/tuberculosis-incidence-uncertainty-intervals?country=~UGA},
}

@Article{McIntosh2019,
  author    = {McIntosh, Avery I. and Jenkins, Helen E. and Horsburgh, C. Robert and Jones-López, Edward C. and Whalen, Christopher C. and Gaeddert, Mary and Marques-Rodrigues, Patricia and Ellner, Jerrold J. and Dietze, Reynaldo and White, Laura F.},
  journal   = {PLOS One},
  title     = {Partitioning the risk of tuberculosis transmission in household contact studies},
  year      = {2019},
  number    = {10},
  volume    = {14},
  publisher = {Public Library of Science San Francisco, CA USA},
}

@Article{Ahmad2011,
  author    = {Ahmad, Suhail},
  journal   = {Clinical and Developmental Immunology},
  title     = {Pathogenesis, immunology, and diagnosis of latent Mycobacterium tuberculosis infection},
  year      = {2011},
  volume    = {2011},
  publisher = {Hindawi},
}

@Article{Esmail2014,
  author    = {Esmail, H. and Barry 3rd, C. E. and Young, D. B. and Wilkinson, R. J.},
  journal   = {Philosophical Transactions of the Royal Society B: Biological Sciences},
  title     = {The ongoing challenge of latent tuberculosis},
  year      = {2014},
  number    = {1645},
  volume    = {369},
  publisher = {The Royal Society},
}

@Article{Hella2017,
  author    = {Hella, Jerry and Morrow, Carl and Mhimbira, Francis and Ginsberg, Samuel and Chitnis, Nakul and Gagneux, Sebastien and Mutayoba, Beatrice and Wood, Robin and Fenner, Lukas},
  journal   = {Journal of Infection},
  title     = {Tuberculosis transmission in public locations in Tanzania: A novel approach to studying airborne disease transmission},
  year      = {2017},
  number    = {3},
  volume    = {75},
  publisher = {Elsevier},
}

@Article{WHO2025,
  author = {{World Health Organization}},
  title  = {Global tuberculosis report 2025},
  year   = {2025},
  url    = {https://iris.who.int/server/api/core/bitstreams/e97dd6f4-b567-4396-8680-717bac6869a9/content},
}

@Article{Melsew2018,
  author    = {Melsew, Y. A. and Doan, T. N. and Gambhir, M. and Cheng, A. C. and McBryde, E. and Trauer, J. M.},
  journal   = {Epidemiology \& Infection},
  title     = {Risk factors for infectiousness of patients with tuberculosis: a systematic review and meta-analysis},
  year      = {2018},
  number    = {3},
  volume    = {146},
  publisher = {Cambridge University Press},
}

@Article{Herriott2025,
  author    = {Herriott, Lara and Capel, Henriette L. and Ellmen, Isaac and Schofield, Nathan and Zhu, Jiayuan and Lambert, Ben and Gavaghan, David and Bouros, Ioana and Creswell, Richard and Gallagher, Kit},
  journal   = {Scientific Reports},
  title     = {EpiGeoPop: a tool for developing spatially accurate country-level epidemiological models},
  year      = {2025},
  number    = {1},
  volume    = {15},
  publisher = {Nature Publishing Group UK London},
}

@Article{Alege2024,
  author    = {Alege, Abiola and Hashmi, Sumbul and Eneogu, Rupert and Meurrens, Vincent and Budts, Anne-Laure and Pedro, Michael and Daniel, Olugbenga and Idogho, Omokhoudu and Ihesie, Austin and Potgieter, Matthys Gerhardus},
  journal   = {Tropical Medicine and Infectious Disease},
  title     = {Effectiveness of using AI-driven hotspot mapping for active case finding of tuberculosis in Southwestern Nigeria},
  year      = {2024},
  number    = {5},
  volume    = {9},
  publisher = {MDPI},
}

@Article{Nsawotebba2025,
  author  = {Nsawotebba, Andrew and Ibanda, Ivan and Mujuni, Dennis and Kabugo, Joel and Adam, Isa and Wekiya, Enock and Orishaba, Philip and Galiwango, John Baptist and Nabadda, Susan and Kasang, Christa and Joloba, Moses and Turyahabwe, Stavia},
  journal = {[Manuscript submitted for publication]},
  title   = {Delays in the Diagnosis and Treatment of Newly Diagnosed Tuberculosis Patients in Kampala, Uganda between July 2021 and June 2022: A cross-sectional study},
  year    = {2025},
}
\begin{appendices}
\newpage
\section{\texttt{CovidSim} and \texttt{Epiabm}}
\label{appendix:covidsim epiabm}
\subsection{Overview}
\label{sec:overview}
\texttt{CovidSim} is a spatial agent-based model, written in C++, in which individuals exist on a grid of square ``cells''; each of these cells is further divided into ``microcells'', with each microcell representing a small part of the geographic region being modelled. The model is set up so that each cell is split into 9 $\times$ 9 microcells, and the width of each microcell corresponds to 
$1/120$ of a degree of latitude. Individuals are allocated to households and ``places'' (that is, schools and workplaces) within these cells, based upon population density data for the region under investigation, and are assigned attributes, such as age and a corresponding 5-year age group (e.g. if an individual is 3 years old, they belong to the 0-4 age group). Individuals also have a current infectious disease state, which corresponds to those typically used in compartmental models of infectious diseases (i.e. Susceptible, Exposed, Infectious, Recovered and Dead/Removed). The infectious compartment is further sub-divided into the following categories: asymptomatic; mild; requiring a GP visit; requiring a hospital visit; requiring an ICU visit; recovering in the ICU but still infectious. This is to account for the fact that an infection from COVID-19 (which the model was originally developed to investigate) can vary in severity from person to person and that not all infectious individuals are symptomatic. The place to which an individual is assigned may be in a different cell to their household. Within each place, individuals are split into ``place groups'', with individuals in the same place group being more likely to interact with each other than individuals in different place groups. They can interact with other individuals within their household, within their place, and via random social meetings which occur with a probability that decreases with distance from their household. It should be noted that the act of people moving is not explicitly modelled; the simulation uses spatial kernels to determine whether an infectious individual in one cell infects a susceptible individual in another cell. The model is initialised with most individuals belonging to the Susceptible compartment and the remainder being Infected. Infections can be seeded in a number of ways (see Section \ref{sec:model initialization} for more details). To determine where the infection spreads, a force of infection is calculated for each susceptible individual based on their interactions with infectious individuals.\par
\texttt{Epiabm} re-implements \texttt{CovidSim} as a more modular version of the original agent-based model and is fully documented; two backends are provided, one in Python and the other in C++. The code is written so that population generation and storage is more object-oriented than it is in \texttt{CovidSim}; although less efficient, the form of the code is more easily adaptable \cite{Gallagher2024}. Further details of how both models work are provided in the following sections.
\subsection{Initialization of the model}
\label{sec:model initialization}
Populations in \texttt{CovidSim} and \texttt{Epiabm} are generated in one of two ways. If a population density file has been specified, both models read this and add people based on the information within the file. From this, individuals can be placed so that the number of people generated matches the expected number listed in the file for each cell and microcell. It is worth mentioning that \texttt{CovidSim} was released with pre-generated configuration files for some regions and countries, while \texttt{Epiabm} uses a separate software package, \texttt{EpiGeoPop}, to generate these files \cite{Herriott2025}. If no such population file is given, the population is generated in random locations based on the number of cells, number of microcells and total population size; this can be subject to constraints such as how many households and how many places there are in each microcell.\par
In both \texttt{CovidSim} and \texttt{Epiabm}, individuals are assigned ages based upon the number of people within their household, using a heuristic algorithm. The sizes of the households are determined before ages are assigned, using a household size distribution which is specified by the user within the parameter file. The algorithm can be summarized as follows:
\begin{itemize}
    \item Determine the type of household given the household size, e.g. if the household consists of two people, the household could contain two younger adults, two older adults, or a single parent and their child, and determine the number of children within the household, if any are present;
    \item If there are children within the household: \begin{itemize}
        \item determine the age of the youngest child, subject to appropriate constraints;
        \item set the age of each subsequent child (if any) to be a Poisson-distributed number larger than the previous child's age, subject to appropriate constraints, with the mean equal to the mean child age gap;
        \item choose the ages of their parents, subject to appropriate constraints;
    \end{itemize} 
    \item If there are children and additional adults beyond the parents within the household, assume they are grandparents and assign them ages such that the age gap between them and the parents is realistic.
\end{itemize}
\texttt{CovidSim} was built with the option to choose place sizes from either a Poisson, log-normal or power-law distribution. Workplaces and schools are generated and distributed in space according to population density files. In the case of schools, children ``pick'' one of the nearest schools to them, with a probability that decreases with distance to the school. In the case of workplaces, workers ``compete'' for vacancies with a probability that decreases with distance. These allocations are weighted with the probability of a person with a given age being in a certain type of place.\par
\texttt{Epiabm} removed the option to choose place sizes from log-normal distributions, instead choosing school sizes from Poisson distributions and workplace sizes from power-law distributions. It should be noted that these are the same distributions typically used for these place types in \texttt{CovidSim}. Otherwise, the two models are similar in this regard.\par
The user specifies the number of initial infections in both models. \texttt{CovidSim} allows the locations of the initial infections to be chosen in a number of different ways:
\begin{itemize}
    \item The probability of initial infections being within a given microcell can be made to be dependent upon population density data, or;
    \item Certain locations can be specified to be the locations of the initial infections, or;
    \item The locations of the initial infections can be chosen at random (that is, from an initial infection location, find the next seeding location by determining the $x$- and $y$-coordinates using a radius and angle generated uniformly; set this subsequent location to be the new initial location and repeat until enough infections have been seeded).
\end{itemize}
The initialization of infections in \texttt{Epiabm} is similar: locations for the initial infections can be specified by the user, otherwise the infections are seeded randomly throughout the spatial domain (that is, the model draws a random sample uniformly from all the people in the population to be the initially infected individuals).
\subsection{Running the model}
\label{sec:running the model}
In \texttt{Epiabm}, once the model has been initialized, the code checks each member of the population one-by-one per time step using multiple functions (one for potential household infections, another for potential infections in places and one for potential transmission between individuals in different cells) to determine which infection events take place and where they occur. For individuals newly infected, they move to the Infected state and the subsequent progression of their infection is randomly decided according to the probabilities of moving from one infection state to another, and the time it takes for this change of status. Additionally, a further function updates the places of the individuals. If non-pharmaceutical interventions have been included, a function determines if any given intervention should be active given the time the policy has been in place and number of infected individuals. If travellers (i.e. individuals not permanently assigned to any specific place) enter or leave the location being modelled, another function determines when they enter the location, which household they reside in during their stay, and how long they stay, for each of those individuals. It is assumed they are infectious throughout their stay. As our model does not make use of non-pharmaceutical interventions and does not introduce or remove travellers, extra details regarding the functions controlling these aspects of the model are not included in this paper; more information can be found in \cite{Herriott2025}. The mechanisms for updating the population and determining infection events and disease progression in the \texttt{CovidSim} model are similar to the \texttt{Epiabm} model unless otherwise stated.\par
Infection events can occur within households, within places, or outside of households and places (i.e. transmission between individuals in different cells). Separate ``sweep'' functions for each of these possibilities determine how many infection events take place. In the case of household infection events, the susceptible people (``infectees'') in the households of every infectious individual (the ``infectors'') are each checked individually. In the case of place infection events, the infectees are members of the infector's place group (e.g. classroom, office, etc). An alternative method is used to determine whether transmission between individuals in different cells occurs: for each cell, the number of infectious individuals is found, and their cumulative infectiousness is used to find a Poisson-distributed number of infections that the infectors in that cell will cause. A list of infectees from the cells other than the ``infector cell'' is then determined by their distance to a randomly chosen infector. \texttt{CovidSim} and \texttt{Epiabm} differ in how distance affects whether somebody is a potential infectee in an infection event between individuals in different cells. \texttt{Epiabm} uses a cut-off distance, beyond which individuals cannot be infected by an infector in another cell; \texttt{CovidSim} calculates the ratio between the spatial kernel of the distance between the potential infector and infectee, and the spatial kernel of the minimum distance between their cells, adding the infectee to the list of potential transmission events if this ratio is greater than a number chosen uniformly at random between 0 and 1. The weighting given by the spatial kernel for distance $d$ is given by the formula
\begin{equation}
    \frac{1}{(1 + \frac{\text{d}}{\text{a}})^\text{b}},
\end{equation}
where the scale parameter, $a$, and shape parameter, $b$,  can be specified by the user. In addition to the cut-off method, this option is also available in \texttt{Epiabm}, where the scale and shape parameters have default values of $a=b=1$.\par
For each of these types of potential transmissions, a force of infection is calculated, and is compared to a number chosen uniformly at random between 0 and 1. This force of infection is based upon the infector's infectiousness and the infectee's susceptibility in that time step and will typically be a number between 0 and 1, corresponding to certain infection. These are each in turn impacted by whether any non-pharmaceutical interventions are in place and whether or not the infector and/or infectee are care home residents. If the number chosen uniformly at random between 0 and 1 is less than the force of infection, the transmission event occurs and the infectee is exposed to the disease, entering a latent state (the infection state is labelled as `Exposed' in \texttt{Epiabm}). For place infection events, if the infector's infectiousness is sufficiently large (that is, at least equal to 1), all infectees become infected. All infected individuals are assigned the Exposed infection state.\par
The within-host sweep checks each member of the population in turn. If the time their infection status was due to change (their ``time of status change'') has been passed, the function updates their infection status and, if their new status is an infectious one, sets their infectiousness. The individual's next infection status, and the time at which they will transition to this status, is then determined. Next, infection statuses and times of status changes are drawn from user-specified cumulative distribution functions. If the person's time of status change has not been passed, and they are in an infectious state, their infectiousness is updated. Infectiousness is based upon a user-specified infectiousness profile that varies over the course of infection.\par
The places of individuals are updated at every time step. Each place is individually checked and the individuals associated with the place are updated. Whether an individual is added to, or removed from, a place will depend on the mean capacity of such places, the average size of place groups within such places (e.g. classroom sizes in schools), the probability distribution governing their size, and the probability of seeing a person of a certain age in that type of place.
\newpage
\section{Model structures for progression from latent TB to active TB}
\label{appendix: latent to active TB}
The following model structures were identified by \citeauthor{Menzies2018} \cite{Menzies2018}.
\begin{figure}[H]
    \centering
    \resizebox{11cm}{!}{
    \begin{tikzpicture}
        \node[draw, fill = green] (healthy) at (-2,0){$S$};
        \node[draw, fill = orange] (latent) at (0,0){$L$};
        \node[draw, fill = red] (sick) at (2,0){$I$};       
        \draw[-stealth] (healthy.east) -- (latent.west) node[midway,above]{$\lambda$};
        \draw[-stealth] (latent.east) -- (sick.west) node[midway,above]{$c$};
    \end{tikzpicture}
    }
    \caption{\raggedright Structure A. Here, $S$ is the susceptible compartment (labelled ``Healthy'' in our model), $L$ is the latent TB compartment, $I$ is the active TB compartment, $\lambda$ is the force of infection for \textit{M. tb}, and $c$ is the rate of progression from latent TB to active TB.}
    \label{fig:structure A}
\end{figure}
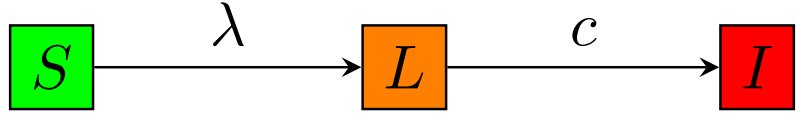
\begin{figure}[H]
    \centering
    \resizebox{11cm}{!}{
    \begin{tikzpicture}[transform shape]
        \node[draw, fill = green] (healthy) at (-3,0){$S$};
        \node[draw, fill = orange, text width = 1.7cm] (latent fast) at (-1,0){\raggedright $L_{f_1},...,L_{f_n}$};
        \node[draw, fill = orange] (latent slow) at (1,0){\raggedright $L_s$};
        \node[draw, fill = red] (sick) at (3,0){\raggedright $I$};   
        \draw[-stealth] (healthy.east) -- (latent fast.west) node[midway,above]{$\lambda$};
        \draw[-stealth] (latent fast.east) -- (latent slow.west) node[midway,above]{$e$};
        \draw[-stealth] (latent slow.east) -- (sick.west) node[midway,above]{$c$};
        \draw[-stealth] (latent fast.north) to[bend left=40] node[midway,above]{$d_1,...,d_n$} (sick.north);        
    \end{tikzpicture}
    }
    \caption{\raggedright Structure B. Here, $L_s$ is the slow latent TB compartment, and $L_{f_1},...,L_{f_n}$ represent $n$ ``tunnel states'' of fast latent TB: if individuals in compartment $L_{f_i},i \in \{1,...,n-1\}$ do not transition to the active TB compartment in a given time step, they progress to state $L_{f_{i+1}}$. $e$ is the rate of progression from fast latent TB to slow latent TB, and $d_1,...,d_n$ are $n$ progression risks: individuals in compartment $L_{f_i},i \in \{1,...,n\}$ transition to the active TB compartment at rate $d_i$. All other symbols have the same meanings as in Figure \ref{fig:structure A}.}
    \label{fig:structure B}
\end{figure}
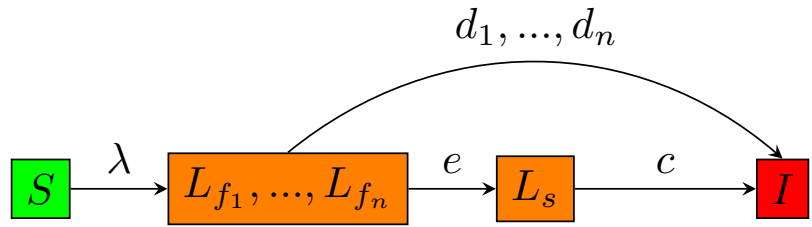
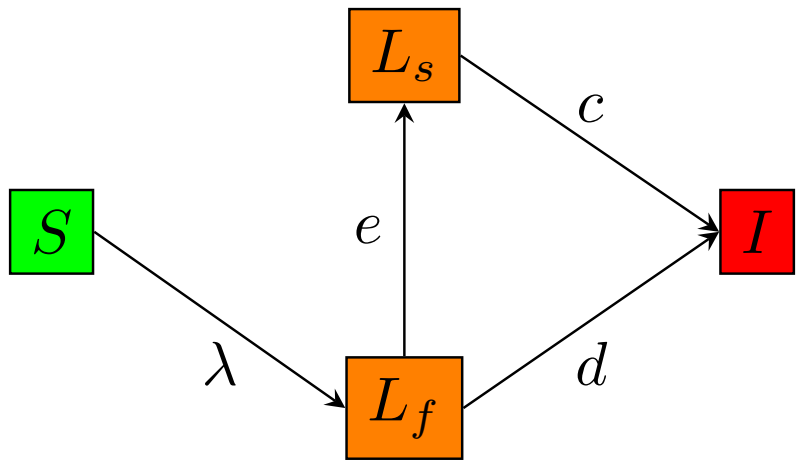
\begin{figure}[H]
    \centering
    \resizebox{11cm}{!}{
    \begin{tikzpicture}
        \node[draw, fill = green] (healthy) at (-2,0){$S$};
        \node[draw, fill = orange] (latent fast) at (0,-1){$L_f$};
        \node[draw, fill = orange] (latent slow) at (0,1){$L_s$};
        \node[draw, fill = red] (sick) at (2,0){$I$};       
        \draw[-stealth] (healthy.east) to node[midway,below]{$\lambda$} (latent fast.west);
        \draw[-stealth] (latent fast.east) to node[midway,below]{$d$} (sick.west);
        \draw[-stealth] (latent fast.north) -- (latent slow.south) node[midway,left]{$e$};
        \draw[-stealth] (latent slow.east) to node[midway,above]{$c$} (sick.west);
    \end{tikzpicture}
    }
    \caption{\raggedright Structure C. Here, $L_f$ is the fast latent TB compartment and $d$ is the rate of progression from fast latent TB to active TB. All other symbols have the same meanings as in Figures \ref{fig:structure A} and \ref{fig:structure B}.}
    \label{fig:structure C}
\end{figure}
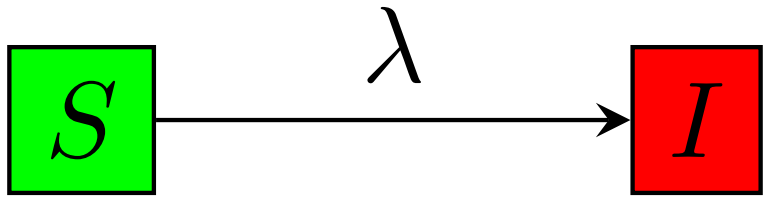
\begin{figure}[H]
    \centering
    \resizebox{11cm}{!}{
    \begin{tikzpicture}
        \node[draw, fill = green] (healthy) at (-1,0){$S$};
        \node[draw, fill = red] (sick) at (1,0){$I$};       
        \draw[-stealth] (healthy.east) -- (sick.west) node[midway,above]{$\lambda$};        
    \end{tikzpicture}
    }
    \caption{\raggedright Structure D. All symbols have the same meanings as in Figures \ref{fig:structure A} to \ref{fig:structure C}.}
    \label{fig:structure D}
\end{figure}
\begin{figure}[H]
    \centering
    \resizebox{11cm}{!}{
    \begin{tikzpicture}
        \node[draw, fill = green] (healthy) at (-2,0){$S$};
        \node[draw, fill = orange] (latent) at (0,0){$L$};
        \node[draw, fill = red] (sick) at (2,0){$I$};       
        \draw[-stealth] (healthy.east) -- (latent.west) node[midway,above]{$\lambda(1-a)$};
        \draw[-stealth] (latent.east) -- (sick.west) node[midway,above]{$c$};
        \draw[-stealth] (healthy.north) to[bend left=40] node[midway,above]{$\lambda a$} (sick.north); 
    \end{tikzpicture}
    }
    \caption{\raggedright Structure E. Here, $a$ is the probability of immediate progression from the susceptible compartment to the active TB compartment after \textit{M. tb} infection. All other symbols have the same meanings as in Figures \ref{fig:structure A} to \ref{fig:structure D}.}
    \label{fig:structure E}
\end{figure}
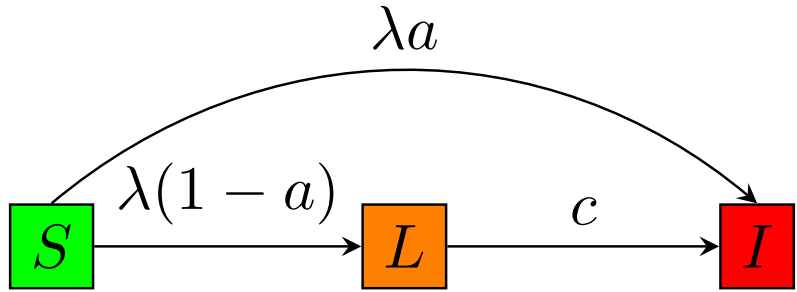
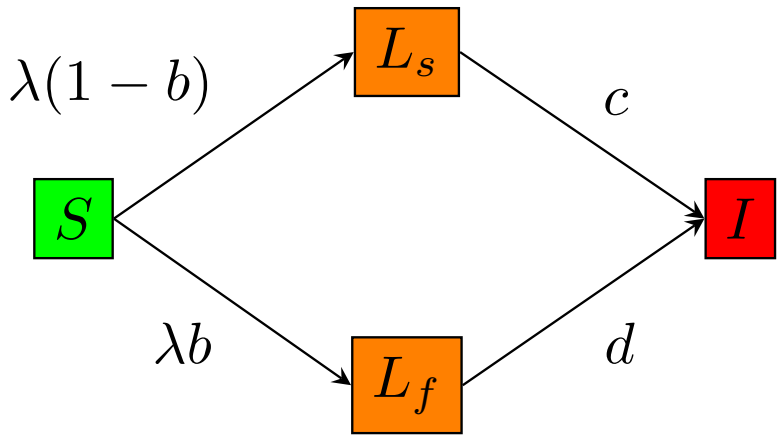
\begin{figure}[H]
    \centering
    \resizebox{11cm}{!}{
    \begin{tikzpicture}
        \node[draw, fill = green] (healthy) at (-2,0){$S$};
        \node[draw, fill = orange] (latent fast) at (0,-1){$L_f$};
        \node[draw, fill = orange] (latent slow) at (0,1){$L_s$};
        \node[draw, fill = red] (sick) at (2,0){$I$};       
        \draw[-stealth] (healthy.east) to node[midway,below left]{$\lambda b$} (latent fast.west);
        \draw[-stealth] (healthy.east) to node[midway,above left]{$\lambda(1-b)$} (latent slow.west);
        \draw[-stealth] (latent fast.east) to node[midway,below right]{$d$} (sick.west);
        \draw[-stealth] (latent slow.east) to node[midway,above right]{$c$} (sick.west);
    \end{tikzpicture}
    }
    \caption{\raggedright Structure F. Here, $b$ is the probability of progression from the susceptible compartment to the fast latent TB compartment after \textit{M. tb} infection. All other symbols have the same meanings as in Figures \ref{fig:structure A} to \ref{fig:structure E}.}
    \label{fig:structure F}
\end{figure}
\begin{figure}[H]
    \centering
    \resizebox{11cm}{!}{
    \begin{tikzpicture}
        \node[draw, fill = green] (healthy) at (-2,0){$S$};
        \node[draw, fill = orange] (latent fast) at (0,-1){$L_f$};
        \node[draw, fill = orange] (latent slow) at (0,1){$L_s$};
        \node[draw, fill = red] (sick) at (2,0){$I$};       
        \draw[-stealth] (healthy.east) to node[midway,below left]{$\lambda(1-a)$} (latent fast.west);
        \draw[-stealth] (latent fast.east) to node[midway,below right]{$d$} (sick.west);
        \draw[-stealth] (latent fast.north) -- (latent slow.south) node[midway,left]{$e$};
        \draw[-stealth] (latent slow.east) to node[midway,below left] {$c$}(sick.west);
        \draw[-stealth] (healthy.north) to[bend left=80] node[midway,above]{$\lambda a$} (sick.north);
    \end{tikzpicture}
    }
    \caption{\raggedright Structure G. All symbols have the same meanings as in Figures \ref{fig:structure A} to \ref{fig:structure F}.}
    \label{fig:structure G}
\end{figure}
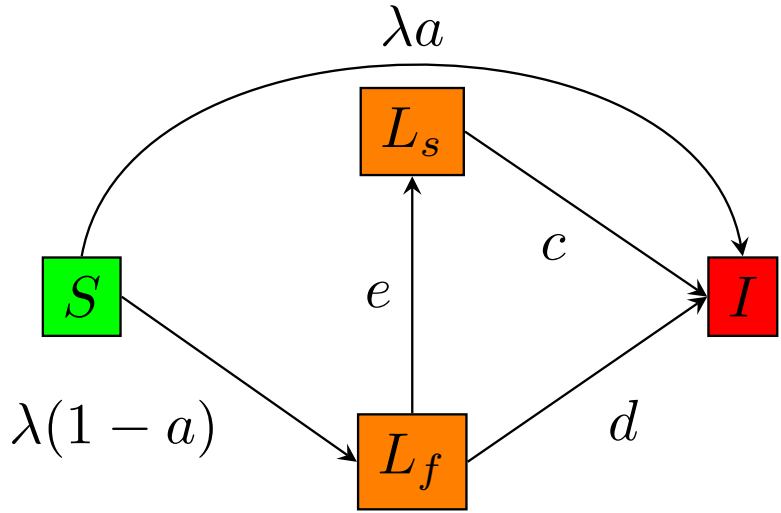
\begin{figure}[H]
    \centering
    \resizebox{11cm}{!}{
    \begin{tikzpicture}
        \node[draw, fill = green] (healthy) at (-2,0){$S$};
        \node[draw, fill = orange] (latent fast) at (0,-1){$L_f$};
        \node[draw, fill = orange] (latent slow) at (0,1){$L_s$};
        \node[draw, fill = red] (sick) at (2,0){$I$};       
        \draw[-stealth] (healthy.east) to node[midway,below left]{$\lambda b$} (latent fast.west);
        \draw[-stealth] (healthy.east) to node[midway,above left]{$\lambda(1-b)$} (latent slow.west);
        \draw[-stealth] (latent fast.east) to node[midway,below right]{$d$} (sick.west);
        \draw[-stealth] (latent slow.south) -- (latent fast.north) node[midway,left]{$f$};
    \end{tikzpicture}
    }
    \caption{\raggedright Structure H. Here, $f$ is the rate of progression from the slow latent TB compartment to the fast latent TB compartment. All other symbols have the same meanings as in Figures \ref{fig:structure A} to \ref{fig:structure G}.}
    \label{fig:structure H}
\end{figure}
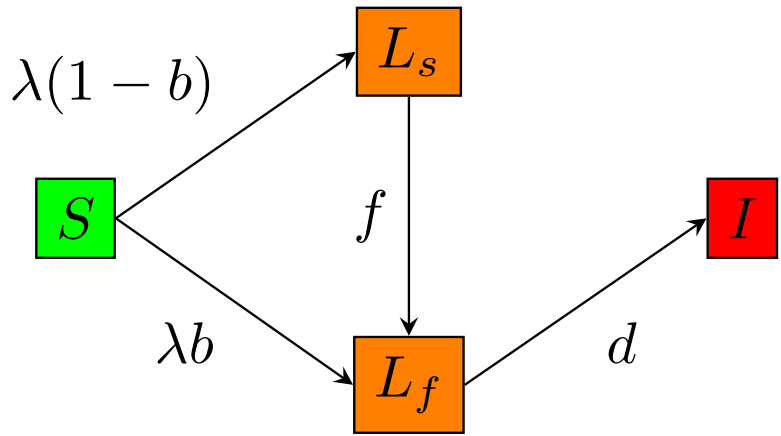
\begin{figure}[H]
    \centering
    \resizebox{11cm}{!}{
    \begin{tikzpicture}
        \node[draw, fill = green] (healthy) at (-2,0){$S$};
        \node[draw, fill = orange] (latent fast) at (0,-1){$L_f$};
        \node[draw, fill = orange] (latent slow) at (0,1){$L_s$};
        \node[draw, fill = red] (sick) at (2,0){$I$};       
        \draw[-stealth] (healthy.east) to node[midway,below]{$\lambda$} (latent fast.west);
        \draw[-stealth] (latent fast.east) to node[midway,below]{$d$} (sick.west);
        \draw[-stealth] (latent fast.north) to[bend right=40] node[midway,right]{$e$} (latent slow.south);
        \draw[-stealth] (latent slow.south) to[bend right=40] node[midway,left]{$f$} (latent fast.north);
    \end{tikzpicture}
    }
    \caption{\raggedright Structure I. All symbols have the same meanings as in Figures \ref{fig:structure A} to \ref{fig:structure H}.}
    \label{fig:structure I}
\end{figure}
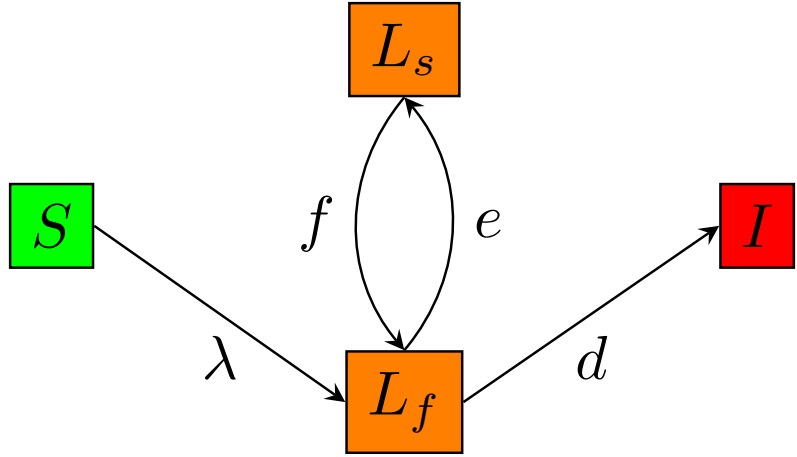
\begin{figure}[H]
    \centering
    \resizebox{11cm}{!}{
    \begin{tikzpicture}[transform shape]
        \node[draw, fill = green] (healthy) at (-3,0){$S$};
        \node[draw, fill = orange, text width = 1.8cm] (latent fast) at (-1,0){\raggedright $L_1,...,L_{n-1}$};
        \node[draw, fill = orange] (latent slow) at (2,0){\raggedright $L_n$};
        \node[draw, fill = red] (sick) at (3,0){\raggedright $I$};   
        \draw[-stealth] (healthy.east) -- (latent fast.west) node[midway,above]{$\lambda$};
        \draw[-stealth] (latent fast.east) -- (latent slow.west) node[midway,above]{$f_{n-1}$};
        \draw[-stealth] (latent slow.east) -- (sick.west) node[midway,above]{$c$};        
    \end{tikzpicture}
    }
    \caption{\raggedright Structure J. Here, $L_1,...,L_n$ represent $n$ sequential latent TB compartments: individuals in compartment $L_{i},i \in \{1,...,n-1\}$ transition to compartment $L_{i+1}$ at rate $f_i$. All other symbols have the same meanings as in Figures \ref{fig:structure A} to \ref{fig:structure I}.}
    \label{fig:structure J}
\end{figure}
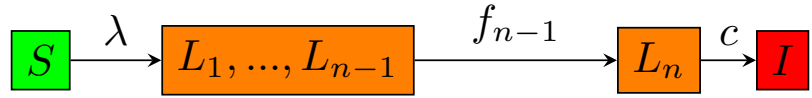
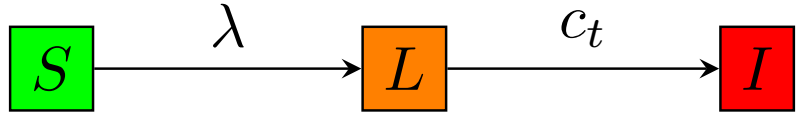
\begin{figure}[H]
    \centering
    \resizebox{11cm}{!}{
    \begin{tikzpicture}
        \node[draw, fill = green] (healthy) at (-2,0){$S$};
        \node[draw, fill = orange] (latent) at (0,0){$L$};
        \node[draw, fill = red] (sick) at (2,0){$I$};       
        \draw[-stealth] (healthy.east) -- (latent.west) node[midway,above]{$\lambda$};
        \draw[-stealth] (latent.east) -- (sick.west) node[midway,above]{$c_t$};
    \end{tikzpicture}
    }
    \caption{\raggedright Structure K. Here, $c_t = x_1 \max(1,t)^{x_2}$ is the rate of progression from latent TB to active TB calculated as a function of an individual's time since infection, $t$, and $x_1$ and $x_2$ are fitted parameter values. All other symbols have the same meanings as in Figures \ref{fig:structure A} to \ref{fig:structure J}.}
    \label{fig:structure K}
\end{figure}
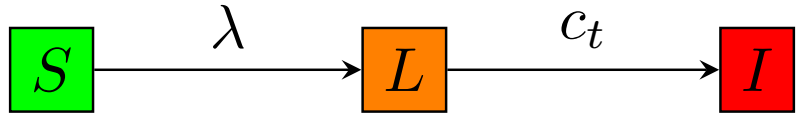
\begin{figure}[H]
    \centering
    \resizebox{11cm}{!}{
    \begin{tikzpicture}
        \node[draw, fill = green] (healthy) at (-2,0){$S$};
        \node[draw, fill = orange] (latent) at (0,0){$L$};
        \node[draw, fill = red] (sick) at (2,0){$I$};       
        \draw[-stealth] (healthy.east) -- (latent.west) node[midway,above]{$\lambda$};
        \draw[-stealth] (latent.east) -- (sick.west) node[midway,above]{$c_t$};
    \end{tikzpicture}
    }
    \caption{\raggedright Structure L. Here, $c_t = x_1 (x_2 + \exp{(-x_3t)})$ is the rate of progression from latent TB to active TB calculated as a function of an individual's time since infection, $t$, and $x_1$, $x_2$ and $x_3$ are fitted parameter values. All other symbols have the same meanings as in Figures \ref{fig:structure A} to \ref{fig:structure K}.}
    \label{fig:structure L}
\end{figure}
\newpage
\clearpage
\section{Parameter values}
\label{appendix:parameter values}
\begin{table}[H]
\centering
\begin{tabular}{|l|l|l|}
\hline
\textbf{Parameter description} & \textbf{Value} & \textbf{Justification}\\
\hline
Population size & 33,612 & \cite{UBOS2019}, see \S\ref{sec:calibrated results}\\
Number of cells & 9 & assumed\\
Number of microcells per cell & 81 & assumed\\
Number of households per microcell & 11 & \cite{UBOS2024}, see \S\ref{sec:calibrated results}\\
Number of places per microcell & 0.17 & \cite{Goyette2014},\cite{MES2017}, see \S\ref{sec:calibrated results}\\
Simulation start time (days) & 0 & default\\
Simulation end time (days) & 2,190 & heuristically chosen\\
%MAY NEED TO REDUCE SIMULATION END TIME IF MODEL TAKES TOO LONG TO RUN
%MAY NEED TO ADJUST INITIAL NUMBER OF PEOPLE IN EACH COMPARTMENT
Initial number of people with latent TB & 1,655 & \cite{Schwalb}, see \S\ref{sec:calibrated results}\\
Initial number of people with active TB & 65 & \cite{Aceng2024}, see \S\ref{sec:calibrated results}\\
Initial number of people receiving treatment & 90 & calibrated, see \S\ref{sec:calibrated results}\\
Initial number of people that could relapse & 425 & calibrated, see \S\ref{sec:calibrated results}\\
\hline
\end{tabular}
\caption{Reference parameter set for initialising the model. Units are given in brackets. Default values were used in the \texttt{basic\_simulation} example on \texttt{Epiabm}'s GitHub web-page: \url{https://github.com/SABS-R3-Epidemiology/epiabm}.}
\label{table:Reference parameter set workflow script}
\end{table}
\begin{table}[H]
\centering
\resizebox{12cm}{!}{
\begin{tabular}{|l|l|l|}
\hline
\textbf{Parameter description} & \textbf{Value} & \textbf{Justification}\\
\hline
Basic reproduction number & 4.3 & \cite{Blower1995}\\
$\mathbb{P}(\text{cavitated disease})$ & 0.4 & \cite{Gadkowski2008},\cite{Zhang2016},\cite{Urbanowski2020}\\
Male cavitated disease probability multiplier & 1.2 & \cite{Balogun2021}\\
Female cavitated disease probability multiplier & 0.8 & \cite{Balogun2021}\\
Mean latent period (days) & 746.863 & \cite{Borgdorff2011}\\
Maximum latent period (years) & 12.8 & \cite{Borgdorff2011}\\
Mean infectious period (days) & 1215.45 & \cite{Tiemersma2011}\\
Maximum infectious period (years) & 10 & \cite{Tiemersma2011}\\
Mean male diagnosis delay (days) & 38.5161 & \cite{Nsawotebba2025}\\
Mean female diagnosis delay (days) & 33.3841 & \cite{Nsawotebba2025}\\
Mean time to stop treatment early (days) & 60 & assumed\\
Treatment period (days) & 180 & \cite{Prats2016}\\
Mean time to relapse (days) & 210 & assumed\\
Maximum time to relapse (days) & 730 & \cite{Prats2016}\\
Number of quantiles per state transition ICDF & 20 & default\\
Time steps per day & 1 & default\\
Maximum household size (people) & 16 & assumed\\
Proportion of females (\%) & 53.0192 & \cite{UBOS2024}\\
Proportion of males (\%) & 46.9808 & \cite{UBOS2024}\\
Place attack rate & 0.506 & \cite{McIntosh2019}\\
Household attack rate & 0.506 & \cite{McIntosh2019}, \cite{Whalen2011}\\
$\mathbb{P}(\text{Individual is vaccinated})$ & 0.83 & \cite{WHOUNICEF2022}\\
Infectiousness decrease due to vaccination (\%) & 74 & \cite{Fromsa2024}\\
Susceptibility decrease due to vaccination (\%) & 20 & \cite{Roy2014}\\
WAIFW matrix stratified by age scalar & 9.9 & calibrated, see \S\ref{sec:calibrated results}\\
WAIFW matrix stratified by gender scalar & 0.44 & calibrated, see \S\ref{sec:calibrated results}\\
Natural death rate (people $\text{day}^{-1}$) & 0.0000133 & \cite{UN2024a}\\
Birth rate (people $\text{day}^{-1}$) & 0.0005 & \cite{UBOS2024}\\
Married proportion (\%) & 56.5 & \cite{UBOS2024}\\
Polygamous marriages (\%) & 13.6 & \cite{UBOS2016}\\
Divorce rate ($\text{year}^{-1}$) & 0.017979 & assumed\\
\hline
\end{tabular}
}
\caption{Reference parameter set for running and updating the model (excluding household age parameters, place parameters, probability distributions, within-host progression probabilities and WAIFW matrix values: see Tables \ref{tab:household ages and places}, \ref{tab:state transition dists}, \ref{tab:within-host}, \ref{tab:household}, \ref{tab:ages}, \ref{tab:WAIFW age} and \ref{tab:WAIFW gender}). Units are given in brackets. Default values were used in the \texttt{basic\_simulation} example on \texttt{Epiabm}'s GitHub web-page: \url{https://github.com/SABS-R3-Epidemiology/epiabm}. Abbreviations: ICDF, inverse cumulative distribution function; WAIFW, Who Acquired Infection From Whom.}
\label{table:Reference parameter set JSON file}
\end{table}
\begin{table}[H]
    \centering
    \resizebox{11.5cm}{!}{
    \begin{tabular}{|l|l|l|}
        \hline
        \textbf{Parameter description} & \textbf{Value} & \textbf{Justification}\\
        \hline
        Mean child age gap (years) & 3 & \cite{UBOS2018}\\
        Minimum adult age (years) & 10 & \cite{UN2024}\\
        Maximum child age (years) & 19 & \cite{UIS2023}\\
        Minimum parent age gap (years) & 10 & \cite{UN2024}\\
        Maximum parent age gap (years) & 59 & assumed\\
        $\mathbb{P}$(one child in a two-person household) & 0.432432 & \cite{UBOS2024},\cite{UNDESA2022}\\
        $\mathbb{P}$(two children in a three-person household) & 0.700855 & \cite{UBOS2024},\cite{UNDESA2022}\\
        $\mathbb{P}$(household consists of one old person) & 0.249 & \cite{UBOS2021}\\
        $\mathbb{P}$(household consists of two old people) & 0.141324 & \cite{UBOS2024},\cite{UNDESA2022}\\
        $\mathbb{P}$(household consists of one young person) & 0.001 & \cite{UBOS2021}\\
        $\mathbb{P}$(household consists of two young people) & 0.000568 & \cite{UBOS2021},\cite{UBOS2024},\cite{UNDESA2022}\\
        $\mathbb{P}$(only child is under five) & 0.274134 & \cite{UBOS2024}\\
        $\mathbb{P}$(youngest of two children is under five) & 0.473119 & \cite{UBOS2024}\\
        $\mathbb{P}$(youngest of 3+ children is under five) & 0.617555 & \cite{UBOS2024}\\
        $\mathbb{P}$(no children in a three-person household) & 0.563674 & \cite{UNDESA2022}\\
        $\mathbb{P}$(one child in a four-person household) & 0.590047 & \cite{UNDESA2022}\\
        $\mathbb{P}$(three children in a five-person household) & 0.409953 & \cite{UNDESA2022}\\
        Maximum partner age gap with older male & 10 & \cite{Kelly2003}\\
        Maximum partner age gap with older female & 0 & \cite{Kelly2003}\\
        Older generation age gap (years) & 26 & \cite{UN2024}\\
        Oldest age for young-person household (years) & 17 & \cite{UBOS2024}\\
        Young and single adult age probability modifier  & 0 & assumed\\
        Minimum age of adults with no children (years) & 55 & \cite{UNDESA2022}\\
        Minimum age for old-person household (years) & 65 & assumed\\
        $\mathbb{P}$(other parent absent in 3+ person house) & 0 & default\\
        Average business size & 124 & \cite{Goyette2014}\\
        Standard deviation of business size & 259 & \cite{Goyette2014}\\
        Average school size & 453 & \cite{UIS2017}\\
        Minimum age of school pupils (years) & 6 & \cite{UIS2023}\\
        Minimum age of teachers in schools (years) & 20 & assumed\\
        Minimum age of workers in workplaces (years) & 15 & \cite{ILO2021}\\
        Maximum age of school pupils (years) & 19 & \cite{UIS2023}\\
        Maximum age of teachers in schools (years) & 64 & assumed\\
        Maximum age of workers in workplaces (years) & 64 & assumed\\
        $\mathbb{P}(\text{Pupil-age male in school})$ & 0.77 & \cite{UIS2017},\cite{UN2024}\\
        $\mathbb{P}(\text{Pupil-age female in school})$ & 0.7903 & \cite{UIS2017},\cite{UN2024}\\
        $\mathbb{P}(\text{20-64 year-old works as a teacher})$ & 0.015 & \cite{UIS2017a},\cite{UN2024}\\
        $\mathbb{P}(\text{15-24 year-old male in workplace})$ & 0.742 & \cite{ILO2021}\\
        $\mathbb{P}(\text{15-24 year-old female in workplace})$ & 0.659 & \cite{ILO2021}\\
        $\mathbb{P}(\text{25-64 year-old male in workplace})$ & 0.9223 & \cite{ILO2021},\cite{UBOS2024}\\
        $\mathbb{P}(\text{25-64 year-old female in workplace})$ & 0.8241 & \cite{ILO2021},\cite{UBOS2024}\\
        Mean classroom size & 40 & \cite{MES2017},\cite{UIS2017}\\
        Mean office size & 10 & \cite{Ragonnet2019}\\
        \hline
    \end{tabular}
    }
    \caption{Reference parameter set for parameters governing household ages and places. Units are given in brackets. Default values were used in the \texttt{basic\_simulation} example on \texttt{Epiabm}'s GitHub web-page: \url{https://github.com/SABS-R3-Epidemiology/epiabm}.}
    \label{tab:household ages and places}
\end{table}
\begin{table}[H]
        \centering
        \begin{tabular}{|l|l|l|l|}
         \hline
         \textbf{Percentile} & \textbf{Exponential} & \textbf{MDD} & \textbf{FDD}\\
         \hline
         $0^{\text{th}}$ & 0 & 0 & 0\\
         $5^{\text{th}}$ & 0.0513 & 0.0628 & 0.0664\\
         $10^{\text{th}}$ & 0.1054 & 0.1264 & 0.1335\\
         $15^{\text{th}}$ & 0.1625 & 0.1915 & 0.202\\
         $20^{\text{th}}$ & 0.2231 & 0.259 & 0.2727\\
         $25^{\text{th}}$ & 0.2877 & 0.3299 & 0.3465\\
         $30^{\text{th}}$ & 0.3567 & 0.4054 & 0.4246\\
         $35^{\text{th}}$ & 0.4308 & 0.4868 & 0.5085\\
         $40^{\text{th}}$ & 0.5108 & 0.5757 & 0.6\\
         $45^{\text{th}}$ & 0.5978 & 0.6745 & 0.7018\\
         $50^{\text{th}}$ & 0.6931 & 0.7862 & 0.8177\\
         $55^{\text{th}}$ & 0.7985 & 0.9161 & 0.9527\\
         $60^{\text{th}}$ & 0.9163 & 1.0722 & 1.1132\\
         $65^{\text{th}}$ & 1.0498 & 1.2655 & 1.3051\\
         $70^{\text{th}}$ & 1.204 & 1.4997 & 1.5286\\
         $75^{\text{th}}$ & 1.3863 & 1.764 & 1.7777\\
         $80^{\text{th}}$ & 1.6094 & 2.0297 & 2.051\\
         $85^{\text{th}}$ & 1.8971 & 2.2637 & 2.3538\\
         $90^{\text{th}}$ & 2.3026 & 2.4797 & 2.6847\\
         $95^{\text{th}}$ & 2.9957 & 2.7268 & 3.0672\\
         $100^{\text{th}}$ & $\frac{\text{max}}{\text{mean}}$ & 94.7656 & 109.3335\\
         \hline
        \end{tabular}
    \caption{Probability distributions for state transitions in reference parameter set for parameter JSON file. All percentiles are given as multiples of the mean. Abbreviations: MDD, male diagnosis delay; FDD, female diagnosis delay. Values for the MDD and FDD distributions were calculated using \citeauthor{Nsawotebba2025} \cite{Nsawotebba2025}. The $100^{\text{th}}$ percentile values for the MDD and FDD distributions were based on the assumption that the maximum diagnosis delay is equal to 10 years.}
    \label{tab:state transition dists}
\end{table}
\begin{table}[H]
    \centering
    \resizebox{12cm}{!}{
    \begin{tabular}{|l|l|l|}
         \hline
         \textbf{Parameter} & \textbf{Value} & \textbf{Justification}\\
         \hline
         $\mathbb{P}$(Latent TB to Active TB) & 0.183 & \cite{Emery2021}\\
         Male multiplier for $\mathbb{P}$(Latent TB to Active TB) & 1.62 & \cite{Gao2017}\\
         Female multiplier for $\mathbb{P}$(Latent TB to Active TB) & 0.38 & \cite{Gao2017}\\
         $\mathbb{P}$(fast TB) & 0.1 & \cite{Ahmad2011}\\
         $\mathbb{P}$(Active TB to TB death) & 0.081 & \cite{WHO2023}\\
         Male multiplier for $\mathbb{P}$(Active TB to TB death) & 1.333 & \cite{JimenezCorona2006},\cite{Fox2016}\\
         Female multiplier for $\mathbb{P}$(Active TB to TB death) & 0.667 & \cite{JimenezCorona2006},\cite{Fox2016}\\
         $\mathbb{P}$(Active TB to Under Treatment) & 0.919 & \cite{WHO2023}\\
         $\mathbb{P}$(Under Treatment to Active TB) & 0.1 & \cite{WHO2024}\\
         $\mathbb{P}$(Under Treatment to Recovered but Could Relapse) & 0.9 & \cite{WHO2024}\\
         $\mathbb{P}$(Recovered but Could Relapse to Active TB) & 0.1 & \cite{Colangeli2018}\\
         \hline
    \end{tabular}
    }
    \caption{Within-host progression probabilities}
    \label{tab:within-host}
\end{table}
\begin{table}
    \centering
    \begin{tabular}{|l|l|l|l|}
        \hline
        \textbf{Household size} & \textbf{HSD}\\
        \hline
        1 & 0.051\\
        2 & 0.074\\
        3 & 0.117\\
        4 & 0.141\\
        5 & 0.27645\\
        6 & 0.176928\\
        7 & 0.094361\\
        8 & 0.043137\\
        9 & 0.017255\\
        10 & 0.006135\\
        11 & 0.001963\\
        12 & 0.000571\\
        13 & 0.000152\\
        14 & 0.000037\\
        15 & 0.000009\\
        16 & 0.000002\\
        \hline
    \end{tabular}
    \caption{Household size in reference parameter set for parameter JSON file. Abbreviation: HSD, household size distribution. Household size distribution values were derived based on \cite{UBOS2024} and \cite{Jarosz2021}.}
    \label{tab:household}
\end{table}
\begin{table}
    \centering
    \begin{tabular}{|l|l|l|l|}
         \hline
         \textbf{Age group} & \textbf{AGD} & \textbf{AID} & \textbf{ACD}\\
         \hline
         0-4 & 0.1477 & 0.000004 & 0.78\\
         5-9 & 0.1425 & 0.000553 & 0.53\\
         10-14 & 0.1324 & 0.075858 & 0.53\\
         15-19 & 0.1161 & 0.924142 & 0.54\\
         20-24 & 0.0963 & 0.999447 & 0.46\\
         25-29 & 0.0795 & 0.999996 & 0.29\\
         30-34 & 0.0631 & 1 & 0.21\\
         35-39 & 0.0528 & 1 & 0.15\\
         40-44 & 0.0423 & 1 & 0.12\\
         45-49 & 0.0319 & 1 & 0.07\\
         50-54 & 0.0275 & 1 & 0.07\\
         55-59 & 0.018 & 1 & 0.05\\
         60-64 & 0.0165 & 1 & 0.05\\
         65-69 & 0.0095 & 1 & 0.08\\
         70-74 & 0.008 & 1 & 0.07\\
         75+ & 0.0159 & 1 & 0.02\\
         \hline
    \end{tabular}
    \caption{Age distributions in reference parameter set for parameter JSON file. Abbreviations: AGD, age group distribution; AID, age infectiousness distribution; ACD, age contact distribution (used for spatial transmissions). Age group distribution values were derived based on \cite{UBOS2024} and rounded to 4 decimal places. The 75+ age group was further split into 75-79 (31.1860\% of the 75+ age group to 4 decimal places); 80-84 (28.7061\%); 85-89 (25.7229\%); 90-94 (10.8811\%) and 95+ (3.5039\%). Age infectiousness distribution values were derived based on \cite{Ragonnet2019}. Age contact distribution values were derived based on \cite{Prem2017}.}
    \label{tab:ages}
\end{table}
\begin{table}
    \centering
    \resizebox{12cm}{!}{
        \begin{tabular}{|r|r|r|r|r|r|r|r|r|r|r|r|r|r|r|r|r|}
         \hline
           & \textbf{1} & \textbf{2} & \textbf{3} & \textbf{4} & \textbf{5} & \textbf{6} & \textbf{7} & \textbf{8} & \textbf{9} & \textbf{10} & \textbf{11} & \textbf{12} & \textbf{13} & \textbf{14} & \textbf{15} & \textbf{16}\\
         \hline
         \textbf{1} & 8.901 & 4.138 & 1.831 & 1.021 & 1.314 & 1.633 & 1.532 & 1.134 & 0.579 & 0.345 & 0.312 & 0.253 & 0.129 & 0.087 & 0.037 & 0.016\\
         \textbf{2} & 4.159 & 23.762 & 4.105 & 1.118 & 0.631 & 1.247 & 1.387 & 1.167 & 0.854 & 0.386 & 0.243 & 0.219 & 0.137 & 0.080 & 0.031 & 0.017\\
         \textbf{3} & 1.414 & 6.262 & 15.817 & 1.862 & 0.786 & 0.620 & 0.784 & 0.821 & 0.703 & 0.411 & 0.219 & 0.118 & 0.067 & 0.063 & 0.032 & 0.017\\
         \textbf{4} & 0.716 & 1.620 & 5.251 & 12.645 & 2.615 & 1.208 & 0.861 & 0.980 & 0.890 & 0.724 & 0.411 & 0.231 & 0.108 & 0.035 & 0.016 & 0.009\\
         \textbf{5} & 0.956 & 0.803 & 0.779 & 4.251 & 5.198 & 2.408 & 1.469 & 1.299 & 0.957 & 0.848 & 0.623 & 0.372 & 0.189 & 0.019 & 0.012 & 0.008\\
         \textbf{6} & 1.469 & 0.878 & 0.453 & 1.332 & 2.824 & 3.311 & 1.989 & 1.592 & 1.305 & 0.958 & 0.836 & 0.476 & 0.241 & 0.017 & 0.005 & 0.006\\
         \textbf{7} & 1.408 & 1.929 & 1.306 & 0.711 & 1.318 & 2.046 & 2.267 & 1.872 & 1.429 & 1.123 & 0.767 & 0.547 & 0.238 & 0.020 & 0.010 & 0.005\\
         \textbf{8} & 1.318 & 1.909 & 1.397 & 1.075 & 0.904 & 1.545 & 1.722 & 2.158 & 1.864 & 1.250 & 0.952 & 0.493 & 0.186 & 0.036 & 0.013 & 0.004\\
         \textbf{9} & 0.791 & 1.322 & 1.217 & 1.165 & 0.988 & 1.317 & 1.563 & 1.687 & 1.913 & 1.513 & 1.142 & 0.480 & 0.239 & 0.027 & 0.007 & 0.006\\
         \textbf{10} & 0.514 & 1.002 & 1.010 & 1.443 & 0.821 & 1.025 & 1.248 & 1.403 & 1.382 & 1.336 & 0.969 & 0.568 & 0.195 & 0.018 & 0.011 & 0.011\\
         \textbf{11} & 0.583 & 0.913 & 1.114 & 1.141 & 0.802 & 1.183 & 1.228 & 1.229 & 1.600 & 1.609 & 1.283 & 0.819 & 0.250 & 0.019 & 0.010 & 0.013\\
         \textbf{12} & 0.988 & 1.288 & 1.052 & 1.029 & 0.709 & 1.031 & 1.219 & 1.065 & 1.228 & 1.033 & 1.073 & 0.815 & 0.358 & 0.038 & 0.015 & 0.022\\
         \textbf{13} & 0.794 & 0.892 & 0.730 & 0.572 & 0.480 & 0.652 & 0.706 & 0.828 & 0.770 & 0.758 & 0.653 & 0.664 & 0.228 & 0.079 & 0.024 & 0.006\\
         \textbf{14} & 0.504 & 0.870 & 0.781 & 0.309 & 0.191 & 0.137 & 0.223 & 0.266 & 0.160 & 0.090 & 0.099 & 0.098 & 0.092 & 0.052 & 0.060 & 0.011\\
         \textbf{15} & 0.227 & 0.746 & 0.573 & 0.406 & 0.067 & 0.102 & 0.095 & 0.169 & 0.069 & 0.133 & 0.108 & 0.098 & 0.093 & 0.081 & 0.027 & 0.053\\
         \textbf{16} & 0.382 & 0.551 & 0.801 & 0.600 & 0.100 & 0.107 & 0.085 & 0.158 & 0.152 & 0.147 & 0.157 & 0.135 & 0.044 & 0.068 & 0.040 & 0.023\\
         \hline
        \end{tabular}
    }
    \caption{Who Acquired Infection From Whom (WAIFW) matrix stratified by age. Age groups: 1, 0-4; 2, 5-9; 3, 10-14; 4, 15-19; 5, 20-24; 6, 25-29; 7, 30-34; 8, 35-39; 9, 40-44; 10, 45-49; 11, 50-54; 12, 55-59; 13, 60-64; 14, 65-69; 15, 70-74; 16, 75+. Values have been taken from \cite{Prem2017} (with justification for doing so coming from \cite{Wallinga2006}) and rounded to 3 decimal places.}
    \label{tab:WAIFW age}
\end{table}
\begin{table}
    \centering
    \begin{tabular}{|r|r|r|}
    \hline
         \diagbox{Infector}{Infectee} & \textbf{Female} & \textbf{Male}\\
          \hline
         \textbf{Female} & 0.37 & 0.3\\
         \textbf{Male} & 0.37 & 0.56\\
         \hline
    \end{tabular}
    \caption{Who Acquired Infection From Whom (WAIFW) matrix stratified by gender. Values have been taken from \cite{Miller2021} (with justification for doing so coming from \cite{Wallinga2006}).}
    \label{tab:WAIFW gender}
\end{table}
\newpage
\clearpage
\section{Justification of log-normal business size distribution assuming Gibrat's law holds}
\label{appendix:lognormal business size}
A business of size $S_t$ at time $t$ (that is, one with $S_t$ employees) will have increased in size by growth rate $\frac{S_t - S_{t-1}}{S_{t-1}} = \varepsilon_t$ from its size at time $t-1$, $S_{t-1}$. In other words, $S_t = S_{t-1}(1 + \varepsilon_t)$. We can iteratively apply this formula to calculate the business sizes at earlier times $t-1, t-2,...,1$ to arrive at the formula $S_t = S_0\underbrace{(1+\varepsilon_1)\cdots(1+\varepsilon_t)}_{t \text{ factors}}$, where $S_0$ is the initial business size. Taking the natural logarithm of both sides gives $\ln S_t = \ln S_0 + \ln (1 + \varepsilon_1) + \cdots +\ln (1+\varepsilon_t)$. If we make infinitely many measurements of the business size between the start time of the business and time $t$, with each measurement coming an infinitesimal amount of time after the previous measurement, then we can assume that the contribution of $\ln S_0$ to $\ln S_t$ will be negligible, and $|\varepsilon_n| < 1$, where $n \in \{1,...,t\}$. By calculating the Taylor expansion of $\ln(1+\varepsilon_n)$, we find it will approximately equal $\varepsilon_n$ for all $n$. Thus, $\ln S_t \approx \varepsilon_1 + \cdots + \varepsilon_t$. If Gibrat's law holds, and the business growth rate is independent of its current size, then we assume that $\varepsilon_n$ is normally distributed with mean $\mu$ and variance $\sigma^2$ for all $n$. In this case, $\ln S_t$ is approximately normally distributed with mean $t\mu$ and variance $t\sigma^2$. Thus, $S_t$ should be approximately log-normally distributed, as expected.
\newpage
\section{Consistency analysis}
\label{sec:consistency analysis epiabm}
A consistency analysis was conducted following the methodology set out by \citeauthor{Hamis2021} \cite{Hamis2021}. This determined the number of simulations required to be sufficient so that the long-term behaviour of quantities such as averages and confidence intervals are consistent. By this, we mean that the differences between the averages and confidence intervals of the outputs of two alternative sets of simulations of the model with the same inputs should have a small statistical significance. Therefore, running this number of simulations should mitigate any intrinsic uncertainty due to the model's stochastic nature. A brief overview of this methodology is described in the following section.\par
The $A$-measure of stochastic superiority (hereafter referred to as the $A$-measure) can be used to compare two distributions, $B$ and $C$ (either continuous or discrete), and check if they are `stochastically equal' with respect to some random variable $X$. In other words, if $X_B$ and $X_C$ denote a random data sample drawn from distributions $B$ and $C$ respectively, then the $A$-measure determines whether $P(X_B > X_C) = P(X_C > X_B)$ \cite{Vargha2000}. The $A$-measure is equal to
\begin{equation}
    A_{12} = P(X_1 > X_2) + 0.5P(X_1 = X_2),
\end{equation}
with $X_1$ and $X_2$ being samples from the two data distributions in question; if the distributions are continuous, $P(X_1 = X_2) = 0$. To estimate the $A$-measure for two discrete distributions with finite sets of data samples drawn from them, where the underlying distributions are not known, an estimate can be taken, hereafter referred to as the $\hat{A}$-measure \cite{Hamis2021} \cite{Alden2013}. For underlying discrete distributions, $B$ and $C$, with $m$ samples $\{b_1,...,b_m\}$ and $n$ samples $\{c_1,...,c_n\}$ (respectively) of random variable $X$ drawn from them, the $\hat{A}$-measure is equal to
\begin{equation}
    \hat{A}_{BC}(X) = \frac{1}{mn} \sum_{i=1}^{m} \sum_{j=1}^{n} \mathbbm{1}_{b_i > c_j} + \frac{\mathbbm{1}_{b_i = c_j}}{2}.
\end{equation}
The $\hat{A}$-measure takes values between 0 and 1: if it equals 0.5, the two distributions are estimated to be `stochastically equal' with respect to $X$; if it is between 0.5 and 1, distribution $B$ is estimated to be `stochastically greater' than distribution $C$ with respect to $X$; if it is between 0 and 0.5, distribution $B$ is estimated to be `stochastically smaller' than distribution $C$ with respect to $X$. As we primarily want to estimate if any two distributions of the same size, $n$, are stochastically equal, it makes sense to scale the $\hat{A}$-measure so that values an equivalent distance in magnitude from 0.5, regardless of whether they are less than 0.5 or greater than 0.5, are scaled to be equivalent. For all values of $n$, the scaled $\hat{A}$-measure, $\underline{\hat{A}}_{1,k'}^n(X),k' \in \{2,...,20\}$ is then calculated for each group of 20 distributions of $n$ samples as follows:
\begin{equation}
\label{eq:scaled A-hat}
    \underline{\hat{A}}_{B,C}^n(X) = \left\{
    \begin{matrix}
    \hat{A}_{B,C}^n(X) & \text{if $\hat{A}_{B,C}^n(X) \ge 0.5$,}\\
    1-\hat{A}_{B,C}^n(X) & \text{if $\hat{A}_{B,C}^n(X) < 0.5$,}
    \end{matrix}
    \right .
\end{equation}
for data samples $B$ and $C$ of size $n$ with regard to random variable $X$. The $\hat{A}$-measure in Equation (\ref{eq:scaled A-hat}) equals:
\begin{equation}
    \hat{A}_{B,C}^n(X) = \frac{1}{n^2}\sum_{i=1}^{n}\sum_{j=1}^{n}H(b_i - c_j),
\end{equation}
where $H(x)$ is the Heaviside function, defined as:
\begin{equation}
    H(x) = \left \{
    \begin{matrix}
        1 & \text{if $x > 0$,}\\
        \frac{1}{2} & \text{if $x = 0$,}\\
        0 & \text{if $x < 0$.}
    \end{matrix}
    \right .
\end{equation}
For each value of $n$ considered, the maximum over $k' \in \{2,...,20\}$ of the scaled $\hat{A}$-measures is determined. The minimum amount of runs recommended by the consistency analysis, $n^*$, is the smallest $n$ such that $\max \limits_{k' \in \{2,...,20\}}(\underline{\hat{A}}_{1,k'}^n(X))$ is below some threshold. Based on the suggestion by \citeauthor{Hamis2021}, this threshold is set equal to 0.56 \cite{Hamis2021}. A scaled $\hat{A}$-measure below this threshold indicates small statistical significance in stochastic differences between samples of a given distribution size.\par
\begin{figure}
    \centering
    \begin{subfigure}{0.7\linewidth}
        \centering
        \includegraphics[width=\linewidth]{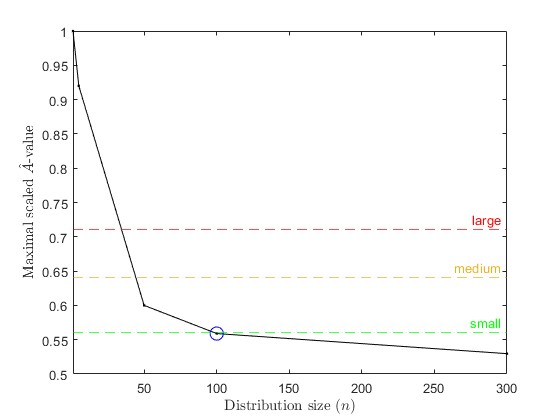}
        \label{fig:ratio consistency analysis}
        \caption{Ratio of male-to-female TB cases.}
    \end{subfigure}
    \hfill
    \begin{subfigure}{0.7\linewidth}
        \centering
        \includegraphics[width=\linewidth]{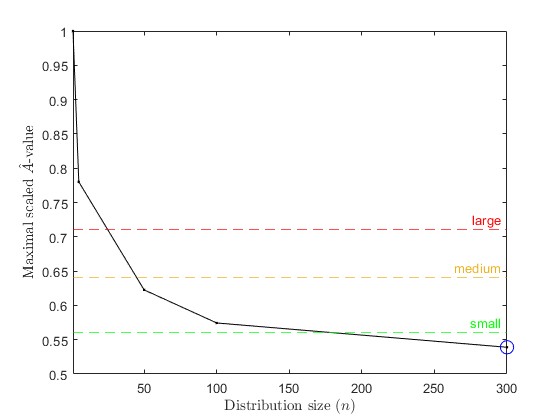}
        \label{fig:total cases consistency analysis}
        \caption{Total number of TB cases.}    
    \end{subfigure}
    \hfill
    \caption[]{Consistency analysis for the \texttt{EpiabmTB} model. The maximal scaled $\hat{A}$-values for various distribution sizes $n$ are shown. The red, yellow and green dashed lines indicate the thresholds below which the scaled $\hat{A}$-values have large, medium and small statistical significance in terms of the stochastic differences between samples of a given distribution size for the random variable being analysed. In the plot for each random variable analysed, the point surrounded by a blue circle is the $n^*$ value for that random variable, that is, the smallest value of $n$ such that the maximal scaled $\hat{A}$-value for the random variable in question is below the threshold for small statistical significance in stochastic differences.}
    \label{fig:consistency analysis}
\end{figure}

A total of $20 \times (1 + 5 + 50 + 100 + 300) = 9,120$ simulations were run to produce the data samples used in the consistency analysis. The outputs of the first 20 runs were grouped into 20 distributions of size 1 for each output. The outputs of the next 100 runs were grouped into 20 distributions of size 5 for each output, and so on: for each $n \in \{1,5,50,100,300\}$ and each output considered, 20 distributions of size $n$ were generated. Two output random variables were analysed: the total number of TB cases at the end of each simulation, and the ratio between male TB cases and female TB cases at the end of each simulation. The consistency analysis suggested $n^* \le 100$ for the ratio, and $n^* \le 300$ for the total number of cases (see Figure \ref{fig:consistency analysis}). From this analysis, we are confident that 300 simulations for both the calibrated results and each counterfactual scenario should mitigate stochastic effects.
\section{Proof of equivalence of the negative binomial distribution and the gamma-Poisson distribution}
\label{appendix:NB gamma-Poisson equivalence}
In this section, we provide a proof of the equivalence of the negative binomial distribution with parameters $r = \frac{\theta}{\rho} > 0$ and $p = \frac{1}{1+\rho} \in [0,1]$, and the Poisson distribution with rate parameter $\lambda \in (0,\infty)$, where $\lambda$ is Gamma-distributed with shape parameter $r = \frac{\theta}{\rho} > 0$ and scale parameter $\rho > 0$. In the following proof:
\begin{itemize}
    \item $f_{\text{Pois}(\lambda)}(k)$ is the probability mass function of the Poisson distribution with rate parameter $\lambda \in (0,\infty)$ evaluated at $k \in \mathbb{N}_0$;
    \item $f_{\Gamma \left (\frac{\theta}{\rho},\rho \right )}(\lambda)$ is the probability density function of the Gamma distribution with shape parameter $\frac{\theta}{\rho} > 0$ and scale parameter $\rho > 0$ evaluated at $\lambda \in (0,\infty)$;
    \item $f_{\Gamma \left (\frac{\theta}{\rho}+k,\frac{\rho}{1+\rho} \right )}(\lambda)$ is the probability density function of the Gamma distribution with shape parameter $\frac{\theta}{\rho} + k > 0$ and scale parameter $\frac{\rho}{1+\rho} > 0$ evaluated at $\lambda \in (0,\infty)$;
    \item $f_{\text{NB} \left (\frac{\theta}{\rho},\frac{1}{1+\rho} \right )}(k)$ is the probability mass function of the negative binomial distribution with parameters $\frac{\theta}{\rho} > 0$ and $\frac{1}{1+\rho} \in [0,1]$ evaluated at $k \in \mathbb{N}_0$;
    \item $\Gamma(z)$ is the Gamma function evaluated at $z > 0$.
\end{itemize}
All other symbols have their standard mathematical interpretation. To prove the two distributions are equivalent, we want to show that, for any input $k$, the probability mass function of the negative binomial distribution evaluated at $k$ is identical to the product of the probability mass function of the Poisson distribution with rate parameter $\lambda$ evaluated at $k$ and the probability density function of the Gamma distribution evaluated at $\lambda$, integrated over all possible values of $\lambda$. We begin the proof by calculating this integral.
\begin{align} 
    &\int_0^\infty f_{\text{Pois}(\lambda)}(k)\times f_{\Gamma \left ( \frac{\theta}{\rho},\rho \right )}(\lambda) d\lambda\\
    = &\int_0^\infty \frac{\lambda^k}{k!}\exp{(-\lambda)} \times \frac{1}{\Gamma \left (\frac{\theta}{\rho} \right )\rho^{\frac{\theta}{\rho}}} \lambda^{\frac{\theta}{\rho}-1} \exp{\left (-\frac{\lambda}{\rho} \right )} d\lambda    
\end{align}
In the above step, we substitute in the probability mass function of the given Poisson distribution and the probability density function of the given Gamma distribution. We can take the constant $\frac{1}{\rho^{\frac{\theta}{\rho}}} \frac{1}{k!\Gamma \left (\frac{\theta}{\rho} \right )}$ outside of the integral and multiply the different powers of $\lambda$ and $e$ together to obtain the following.
\begin{align}
    &\int_0^\infty \frac{\lambda^k}{k!}\exp{(-\lambda)} \times \frac{1}{\Gamma \left (\frac{\theta}{\rho} \right )\rho^{\frac{\theta}{\rho}}} \lambda^{\frac{\theta}{\rho}-1} \exp{\left (-\frac{\lambda}{\rho} \right )} d\lambda\\
    = &\frac{1}{\rho^{\frac{\theta}{\rho}}}\frac{1}{k!\Gamma \left (\frac{\theta}{\rho} \right )} \int_0^\infty \lambda^{\frac{\theta}{\rho} + k - 1} \exp{\left (-\lambda \frac{1 + \rho}{\rho} \right )} d\lambda   
\end{align}
The probability density function of the Gamma distribution with shape parameter $\frac{\theta}{\rho}+k$ and scale parameter $\frac{\rho}{1+\rho}$ evaluated at $\lambda$ is equal to $\frac{1}{\Gamma \left (\frac{\theta}{\rho} + k \right ) \left (\frac{\rho}{1+\rho} \right )^{\frac{\theta}{\rho}+k}}\lambda^{\frac{\theta}{\rho}+k-1}\exp{\left (-\lambda\frac{1+\rho}{\rho} \right )}$. We can use this fact to simplify the expression as follows.
\begin{align} 
    &\frac{1}{\rho^{\frac{\theta}{\rho}}}\frac{1}{k!\Gamma \left (\frac{\theta}{\rho} \right )} \int_0^\infty \lambda^{\frac{\theta}{\rho} + k - 1} \exp{\left (-\lambda \frac{1 + \rho}{\rho} \right )} d\lambda\\
    = &\frac{1}{\rho^{\frac{\theta}{\rho}}}\frac{1}{k!\Gamma \left (\frac{\theta}{\rho} \right )}\Gamma \left (\frac{\theta}{\rho}+k \right ) \left (\frac{\rho}{1 + \rho} \right )^{\frac{\theta}{\rho} + k} \int_0^\infty f_{\Gamma \left (\frac{\theta}{\rho} + k,\frac{\rho}{1 + \rho} \right )}(\lambda) d\lambda
\end{align}
A probability density function must integrate to 1 over its support, so we can determine that the integral is equal to 1 and subsequently remove it, then simplify the expression to arrive at the following.
\begin{align}     
    &\frac{1}{\rho^{\frac{\theta}{\rho}}}\frac{1}{k!\Gamma \left (\frac{\theta}{\rho} \right )}\Gamma \left (\frac{\theta}{\rho}+k \right ) \left (\frac{\rho}{1 + \rho} \right )^{\frac{\theta}{\rho} + k} \int_0^\infty f_{\Gamma \left (\frac{\theta}{\rho} + k,\frac{\rho}{1 + \rho} \right )}(\lambda) d\lambda\\
    = &\frac{\Gamma \left (\frac{\theta}{\rho} + k \right )}{k! \Gamma \left (\frac{\theta}{\rho} \right )} \left (\frac{\rho}{1 + \rho} \right )^k \left (\frac{1}{1 + \rho} \right )^{\frac{\theta}{\rho}}
\end{align}
The function we are left with is equal to the probability mass function of the negative binomial distribution we were aiming to prove was equivalent to the integral stated at the start of the proof (as $\frac{\rho}{1+\rho} = 1 - \frac{1}{1+\rho}$).
\begin{align}    
    &\frac{\Gamma \left (\frac{\theta}{\rho} + k \right )}{k! \Gamma \left (\frac{\theta}{\rho} \right )} \left (\frac{\rho}{1 + \rho} \right )^k \left (\frac{1}{1 + \rho} \right )^{\frac{\theta}{\rho}}\\
    = &f_{\text{NB} \left (\frac{\theta}{\rho},\frac{1}{1 + \rho} \right )}(k)
\end{align}
Hence, our proof is complete.
\newpage
\clearpage
\section{Super-spreaders by stratification}
\label{appendix:stratified superspreaders}
\begin{figure}[H]
    \centering
    \includegraphics[width=\linewidth]{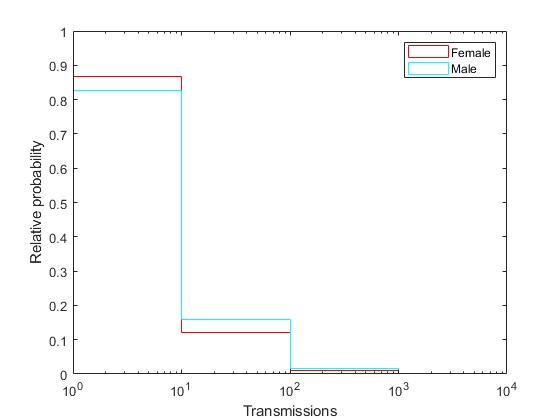}
    \caption{Probability distribution of the number of transmissions per infector (who infected at least one individual), stratified by gender, averaged across the 300 simulations. The red histogram bars indicate the average relative probability (that is, the number of elements in each bin relative to the total number of elements in the input data) of the number of transmissions per female infector, and the cyan histogram bars indicate the average relative probability of the number of transmissions per male infector. Note the $x$-axis is on a base-10 logarithmic scale.}
    \label{fig:transmissions gender}
\end{figure}
\begin{figure}
    \centering
    \includegraphics[width=\linewidth]{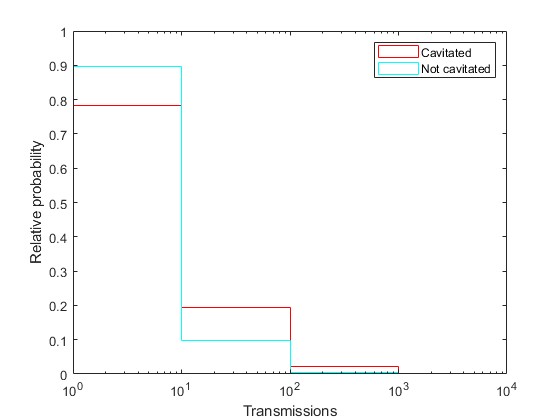}
    \caption{Probability distribution of the number of transmissions per infector (who infected at least one individual), stratified by cavitation status, averaged across the 300 simulations. The red histogram bars indicate the average relative probability (that is, the number of elements in each bin relative to the total number of elements in the input data) of the number of transmissions per infector with cavitary TB, and the cyan histogram bars indicate the average relative probability of the number of transmissions per infector with non-cavitary TB. Note the $x$-axis is on a base-10 logarithmic scale.}
    \label{fig:transmissions cavitation}
\end{figure}
\begin{figure}
    \centering
    \includegraphics[width=\linewidth]{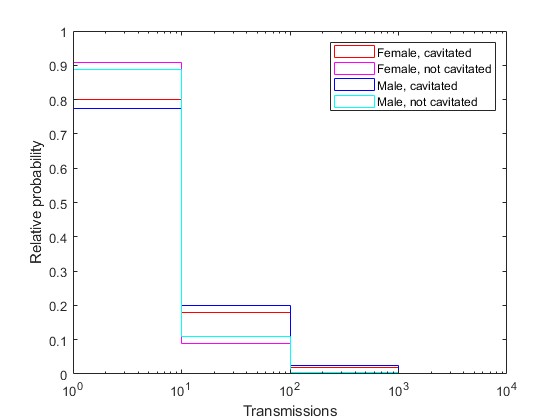}
    \caption{Probability distribution of the number of transmissions per infector (who infected at least one individual), stratified by both gender and cavitation status, averaged across the 300 simulations. The red histogram bars indicate the average relative probability (that is, the number of elements in each bin relative to the total number of elements in the input data) of the number of transmissions per female infector with cavitary TB, the magenta histogram bars indicate the average relative probability of the number of transmissions per female infector with non-cavitary TB, the blue histogram bars indicate the average relative probability of the number of transmissions per male infector with cavitary TB, and the cyan histogram bars indicate the average relative probability of the number of transmissions per male infector with non-cavitary TB. Note the $x$-axis is on a base-10 logarithmic scale.}
    \label{fig:transmissions cavitated}
\end{figure}
\end{appendices}
\end{document}